\newif\ifShowKeys
\ShowKeystrue
\documentclass[11pt,a4paper]{article} 
\usepackage[height=8.8in,width=6.45in]{geometry}

\usepackage[hidelinks]{hyperref}

\usepackage{tocloft}
\usepackage{amsmath, amssymb,amsthm}
\usepackage{amsthm}
\numberwithin{equation}{section}

\newcommand{\rf}[1]{(\ref{#1})}

\usepackage{bm,environ,mathrsfs,array,arydshln}
\usepackage{booktabs,float,slashed,hyperref}
\usepackage[mathcal]{euscript}
\usepackage{tensor} 						% Ratcliffe package to write tensors
\usepackage{mathabx}
\usepackage{simpler-wick}

\usepackage{aurical}
\usepackage[T1]{fontenc}

\usepackage[nodayofweek]{date time}
\usepackage{graphicx,epsfig,epic}
\usepackage{tikz} 
\usetikzlibrary{decorations.pathreplacing,decorations.markings,snakes}
\tikzset{middlearrow/.style={decoration={markings, mark= at position 0.5 with {\arrow{#1}} ,
}, postaction={decorate}}}

\usepackage{framed}						% for shaded equations \begin{shaded}...\end{shaded}
\definecolor{shadecolor}{rgb}{0.95,0.95,0.97}

\usepackage{comment}

\allowdisplaybreaks

\newcommand{\bs}{\begin{shaded}}
\newcommand{\es}{\end{shaded}}
\def\ba#1\ea{\begin{align}#1\end{align}}		% very clever way to bypass the known problem...
\newcommand{\be}{\begin{equation}}
\newcommand{\ee}{\end{equation}}
\newcommand{\mc}{\mathcal }

\newcommand{\la}{\label}
\newcommand{\eps}{\varepsilon}

\newcommand{\lp}{\notag \\ & }

\DeclareMathOperator{\tr}{\text{tr}}

\newcommand{\ie}{\textit{i.e.} }

\makeatletter
\DeclareFontFamily{OMX}{MnSymbolE}{}
\DeclareSymbolFont{MnLargeSymbols}{OMX}{MnSymbolE}{m}{n}
\SetSymbolFont{MnLargeSymbols}{bold}{OMX}{MnSymbolE}{b}{n}
\DeclareFontShape{OMX}{MnSymbolE}{m}{n}{
<-6>  MnSymbolE5
   <6-7>  MnSymbolE6
   <7-8>  MnSymbolE7
   <8-9>  MnSymbolE8
   <9-10> MnSymbolE9
  <10-12> MnSymbolE10
  <12->   MnSymbolE12
}{}
\DeclareFontShape{OMX}{MnSymbolE}{b}{n}{
<-6>  MnSymbolE-Bold5
   <6-7>  MnSymbolE-Bold6
   <7-8>  MnSymbolE-Bold7
   <8-9>  MnSymbolE-Bold8
   <9-10> MnSymbolE-Bold9
  <10-12> MnSymbolE-Bold10
  <12->   MnSymbolE-Bold12
}{}

\let\llangle\@undefined
\let\rrangle\@undefined
\DeclareMathDelimiter{\llangle}{\mathopen}%
 {MnLargeSymbols}{'164}{MnLargeSymbols}{'164}
\DeclareMathDelimiter{\rrangle}{\mathclose}%
 {MnLargeSymbols}{'171}{MnLargeSymbols}{'171}
\makeatother

\catcode`,\active

\catcode`\,12

\def\XXint#1#2#3{{\setbox0=\hbox{$#1{#2#3}{\int}$}
     \vcenter{\hbox{$#2#3$}}\kern-.5\wd0}}

\newcommand{\gs}{g_{\text{s}}}

\newcommand{\vev}[1]{\left\langle  #1 \right\rangle}

\newcommand{\WSYM}{\vev{\mc W}^{\rm SYM}}

\def \N {{\cal N}}

\begin{document}

\begin{titlepage}

%\date{\currenttime}
%\begin{flushright}\boxed{\small{\tt \today \ \ - \ \  \currenttime }}\end{flushright}

\begin{tabbing}
\hspace*{11.5cm} \=  \kill % set the tabbings
\>  {Imperial-TP-AT-2021-01} % additional information:
  \\
\> % none
\end{tabbing}

\vspace*{15mm}
\begin{center}
{\LARGE\sc   $1/N$ expansion of  circular  Wilson loop}\vskip 5pt
{\LARGE\sc   in    $\mc N=2$ superconformal  $SU(N)\times SU(N)$ quiver}

\vspace*{10mm}

{\Large M. Beccaria${}^{\,a}$  and   A.A. Tseytlin${}^{\,b,}$\footnote{\ Also at the Institute of Theoretical and Mathematical Physics, MSU and Lebedev Institute, Moscow. %\\\hspace*{15pt} \ tseytlin@imperial.ac.uk}
}
}

\vspace*{4mm}
	
${}^a$ Universit\`a del Salento, Dipartimento di Matematica e Fisica \textit{Ennio De Giorgi},\\ 
		and I.N.F.N. - sezione di Lecce, Via Arnesano, I-73100 Lecce, Italy
			\vskip 0.3cm
${}^b$ Blackett Laboratory, Imperial College London SW7 2AZ, U.K.
			\vskip 0.3cm
			
\vskip 0.2cm
	{\small
		E-mail:
		\texttt{matteo.beccaria@le.infn.it, \ \ tseytlin@imperial.ac.uk}
	}
\vspace*{0.8cm}
\end{center}

\begin{abstract}  
Localization  approach  to  $\N=2$ superconformal   $SU(N) \times SU(N)$  quiver theory 
  leads to a  non-Gaussian two-matrix model  representation for 
the expectation  value of  BPS   circular $SU(N)$  Wilson loop $\vev{\mc W}$. We study the   subleading  $1/N^2$
 term    in  the large $N$  expansion of $\vev{\mc W}$  at   weak and strong  coupling. 
    We concentrate  on the case 
 of the symmetric quiver  with equal  gauge couplings  which is equivalent to the  $\mathbb Z_{2}$
 orbifold  of  the   $SU(2N)$  $\N=4$ SYM theory. This orbifold gauge theory  should be  dual to type IIB   superstring  in 
 ${\rm AdS}_5\times (S^{5}/\mathbb Z_{2})$. We present a    string theory argument  suggesting 
  that  the    $1/N^2$   term   
 in $\vev{\mc W}$   in the orbifold theory 
  should have the same  strong-coupling   asymptotics $ \lambda^{3/2}$  as in  the $\N=4$ SYM case. 
  We support  this prediction  on the gauge theory side by a  numerical study  of the localization matrix model.
  % representation for  $\vev{\mc W}$. 
  We also  find  a   relation   between   the $1/N^2$  term  in the Wilson loop expectation value and  the  derivative  of the 
  free energy of the orbifold   gauge  theory  on  4-sphere.
  
  %%%%%%%%%%%%%%%%%%
\end{abstract}
\vskip 0.5cm
	{
		%Keywords: {\sc insert here keywords}
	}
\end{titlepage}

\iffalse   FOR SUBMISSION:

Localization  approach  to  $\cal N=2$ superconformal   $SU(N) \times SU(N)$  quiver theory 
  leads to a  non-Gaussian two-matrix model  representation for 
the expectation  value of   the BPS   circular $SU(N)$  Wilson loop 
$\langle  W\rangle $. We study the   subleading $1/N^2$
 term   in   the large $N$  expansion of $\langle  W\rangle $    at   weak and strong  coupling. 
    We concentrate  on the case 
 of the symmetric quiver  with equal  gauge couplings  which is equivalent to the  $\mathbb Z_2$
 orbifold  of  the    $SU(2N)$  $\N=4$ SYM theory. This orbifold gauge theory  should be  dual to  type IIB   superstring  in 
 ${\rm AdS}_5\times (S^{5}/\mathbb Z_{2})$. We present a    string theory argument  suggesting 
  that  the    $1/N^2$   term   in $\langle  W\rangle $
    in the orbifold theory 
  should have the same  strong-coupling   asymptotics $ \lambda^{3/2}$  as in  the $\N=4$ SYM case. 
  We confirm  this prediction  on the gauge theory side by a  numerical study  of the localization matrix model.
  % representation for  $\vev{\mc W}$. 
  We also  find  a   relation   between   the $1/N^2$  term  in the Wilson loop expectation value and  the derivative of the 
  free energy of the orbifold   gauge  theory  on  4-sphere.

\fi

\tableofcontents
\vspace{1cm}

\def \foot{\footnote}\def \ci{cite}\def \l {\lambda}\def \iffa {\iffalse}
\def \RR {{R}} \def \ov {\over }\def \a  {\alpha} \def \ha {{1\ov 2}}
\def \ed {\end{document}}
\def \bi{\bibitem}
\def \lab {\label}
\def \l {\lambda}
\def\foot{\footnote}
\def \adss {${\rm AdS}_5 \times S^5~$ }
\def \ov {\over}
\def \tr {{\rm tr}}
\def \ha {{1 \over 2}}
\def \td {\tilde}
\def \ci{\cite}
\def \bi {\bibitem}
\def \N  {{\cal N}}
\def \aa  {{\rm a}}
\def \te {\textstyle} 
\def \Z {{\cal Z}}
\def \RR {{L}}
\def \aa {{\rm a}} \def \aai {{\aa i}}  \def \aaj {{\aa j}}
\def \eps {\epsilon}\def \rr {{\rm r}}

\setcounter{footnote}{0}

\section{Introduction and summary}

Supersymmetric Wilson loop      operators  provide an important class of observables 
that shed  light on the intricate structure of weak-strong coupling interpolation   in  the context of  AdS/CFT duality. 
In special cases with extended   supersymmetry  the localization  method  \cite{Pestun:2016zxk}  allows  one to  represent the expectation value of a  supersymmetric loop   in terms of a   matrix model integral. 

Here we  will consider a  particular  $\mc N=2$ supersymmetric  gauge theory   which is the $SU(N)\times SU(N)$
quiver with two  bi-fundamental hypermultiplets  %(for a background and references see    
\ci{Rey:2010ry,Passerini:2011fe,Zarembo:2020tpf,Ouyang:2020hwd}. 
In the ``symmetric'' case when   the two  't Hooft  couplings $\lambda_{1}$, $\lambda_{2}$ are equal
this  theory  %has  an additional $\mathbb Z_{2}$ symmetry  and
 is equivalent  to the $\mathbb Z_{2}$ orbifold of  the  $SU(2N)$  $\mc N=4$  SYM  theory  \cite{Lawrence:1998ja}. 
The orbifold theory  has the same planar diagrams as the parent $\mc N=4$ SYM theory \cite{Bershadsky:1998cb},  i.e. 
the two are closely related  at large $N$.  The dual string theory    should be  the  corresponding  orbifold of the 
${\rm AdS}_5\times S^{5}$  superstring, i.e. type IIB   string  on 
 ${\rm AdS}_5\times (S^{5}/\mathbb Z_{2})$ 
 %v2
 \cite{Kachru:1998ys,Gadde:2009dj}.\foot{$\mathbb{Z}_{2}$ acts by flipping 4  of the 6  embedding coordinates of the 5-sphere, reflecting the 2+4 split of the $\mc N = 4$ SYM 
scalars between the vector  multiplets and the hypermultiplets of the  $\mc N = 2$  theory.}

For each of the two $SU(N)$  factors  of the quiver theory % 
one may   define  $\frac{1}{2}$-BPS  circular Wilson loops coupled to  the corresponding  gauge and scalar fields
%$A_{I, \mu}$, $\Phi_{I}$ 
 ($\aa=1,2$)
\be\la{0}
\mc W_\aa = \tr\, \mc P\,\exp \Big[
\oint ds\,(i\,\dot x^{\mu}\,A_{\mu\,\aa}+|\dot x|\,\Phi_\aa) \Big] \ , 
\ee
where we  choose not to  include the  $1/N$  factor in front of the trace. %in the definition of 
For the orbifold theory  their  (normalized) expectation values  are equal 
\be \la{00}
\langle{\mc W_1} \rangle= \langle{\mc W_2}\rangle\equiv \vev{\mc W}^{\rm orb}\ , \ee
 and at large $N$
coincide   \cite{Zarembo:2020tpf} 
with the  famous  $SU(N)$    $\mc N=4$  SYM result  \cite{Erickson:2000af,Drukker:2000rr,Pestun:2007rz}
\ba
\la{1.1}  %\vev{\mc W_{1}} = \vev{\mc W_{2}}\equiv \vev{\mc W}^{\rm orb} \ , \qquad \qquad 
 &\vev{\mc W}^{\rm orb}_{N\to\infty} =  % \stackrel{N\to\infty}{=}
\vev{\mc W}^{\rm SYM}_{N\to\infty} =  \vev{\mc W}_0 \ , \qquad  \notag \\
&\vev{\mc W}_0 = %\stackrel{N\to\infty}{=} 
\frac{2\,N}{\sqrt\lambda}\,I_{1}(\sqrt\lambda) \stackrel{\lambda\gg 1}{=} \sqrt\frac{2}{\pi}\,N\,\lambda^{-3/4}\,e^{\sqrt\lambda}
\Big[1+\mc O\Big(\frac{1}{\sqrt\lambda}\Big)\Big]  \ . %+\cdots \ . 
\ea
%where the exponential term is related to the minimal area law in ${\rm AdS}_5$.
For general $N$ the expression  for $\vev{\mc W}^{\rm orb}$ is given   by  a special non-Gaussian 
  matrix model integral  following from  the  localization approach \cite{Pestun:2007rz}.  
  In contrast to the $\mc N = 4$    SYM    case where the corresponding 
  matrix model is Gaussian leading to the closed   expression   \cite{Drukker:2000rr,Pestun:2007rz}
  \be
\la{1.8}
\vev{\mc W}^{\rm SYM} = e^{\frac{\lambda}{8N}(1-\frac{1}{N})}L^{(1)}_{N-1}\Big(-\frac{\lambda}{4N}\Big) = N\,e^{\sqrt\lambda}\sum_{p=0}^{\infty}
\frac{\sqrt{2}}{96^{p}\sqrt\pi \, p!}\frac{\lambda^{\frac{6p-3}{4}}}{N^{2p}}\Big[1+\mc O\Big(\frac{1}{\sqrt\lambda}\Big)\Big] \ , 
\ee
  working  out the $1/N$ expansion of  $\vev{\mc W}^{\rm orb}$   turns out to be a non-trivial  problem. 
Below we will   address the question    about the   structure of  the 
 $\l$-dependent coefficients  in  the  
$1/N$ expansion  of    $\vev{\mc W}^{\rm orb}$  by considering  separately the   small and  large $\l$  limits.

%Beyond large $N$, the structure of the higher genus corrections to (\ref{1.1}) is a non-trivial issue, despite the available matrix model
%formulation of the quiver by localization \cite{Pestun:2007rz}. 
On the  dual string theory  side, the $1/N^2$   expansion is the   genus expansion, 
and the $\l$ dependence  of the $1/N^{2p}$  coefficient  in the analog of \rf{1.8}
  corresponds to the string tension dependence 
of the partition function with world surface  of topology of  a disc with $p$  handles. 
%In general, the genus expansion of the Wilson loop at strong coupling
%is an important probe for string perturbation theory. 

As discussed recently  in \cite{Giombi:2020mhz}, the strong coupling expansions
 of the  $\frac{1}{2}$-BPS circular Wilson loops in $\mc N=4$ SYM and ABJM gauge theories
with string duals defined on  ${\rm AdS}_5\times S^{5}$ and ${\rm AdS}_{4}\times CP^{3}$ have a remarkably 
similar structure. %  at the first subleading (one-loop) order of the expansion around the minimal surface. 
The string counterpart of the  dominant  at large $N$  ($p=0$)  term in (\ref{1.1}),\rf{1.8}  is the open-string partition function 
on the disk    which contains an overall factor of the inverse  closed  string coupling $\gs$ 
%with Euler number $\chi=1-2p =1$) 
%%%%%%%%%%%%%%%%%%%%%%%%%
\be\la{1.5}
\vev{\mc W}_0  = % \frac{1}{2\pi}\,
%v4
\frac{\sqrt T}{\sqrt{2 \pi}\,  \gs}\,e^{2\pi T}e^{-\bar\Gamma }\Big[1 + \mc O\Big( T^{-1} \Big)\Big]\ .
\ee
In the  $SU(N)$  $\mc N=4$ SYM   case   % string coupling and tension are 
\be\la{1.6}
\gs = \frac{g^{2}_{\rm YM}}{4\pi} = \frac{\lambda}{4\pi N},\qquad \qquad T = { \RR^2 \ov 2\pi \a'}=\frac{\sqrt\lambda}{2\pi}\ , \qquad \ \ \ 
\bar\Gamma =\frac{1}{2}\log(2\pi) \ , 
\ee
so that the leading  term in \rf{1.5} is the same as in \rf{1.1}. 

In general, 
%$\bar\Gamma_{1}$ is a finite numerical constant coming from the ratio of one-loop determinants of string fluctuations near the minimal surface(and being $\frac{1}{2}\log(2\pi)$ in SYM).
the presence of the universal 
$\sqrt T$ prefactor in \rf{1.5}  follows  from  the structure of   the  1-loop    fluctuation 
determinants \cite{Drukker:2000ep} appearing in the string partition function expanded near the AdS$_2$ minimal surface
 (corresponding to the circular Wilson loop). 
In the  case of genus $p$ surface 
  the   UV divergent part of the one-loop effective action $\Gamma  =\ha  \sum_i  \log \det \Delta_i $  reads \cite{Giombi:2020mhz}
\be\la{1.7} 
\Gamma  = -\zeta_{\rm tot}(0)\log(\RR\Lambda)+\bar\Gamma\ ,\qquad\qquad  \zeta_{\rm tot}(0) = \chi=1-2p  \ , 
\ee
where $\Lambda$ is 2d  cutoff, $\RR$  is the AdS radius ($T= {\RR^2\ov 2 \pi\a'}$) and  the 
 $\zeta_{\rm tot}(0)$  coefficient  turns out  to be  equal to the Euler number of the surface. The 2d UV  divergence    should  be canceled  
 by a universal superstring measure contribution $\log(\sqrt{\alpha'}\Lambda)$ involving only the string scale and not the AdS radius.
Then the finite part of $\Gamma$    depends on $T$   through the term
%v4
 $ -\chi\log\frac{\RR}{\sqrt{\alpha'}}
 = - \chi \log \sqrt {T}$  and thus  the string partition function on a genus $p$ surface is proportional to $ e^{-\Gamma_{\rm fin}} \sim 
 (\sqrt T)^{\chi}$, i.e. 
\be\la{18}
\vev{\mc W} =\sum_{p=0}^{\infty} \vev{\mc W}_p =  e^{2\pi\,T}\sum_{p=0}^{\infty}c_{p}\,\Big(\frac{\gs}{\sqrt T}\Big)^{2p-1}\Big[1+\mc O\Big(T^{-1}\Big)\Big] \ .
\ee
Written in terms of $N$ and $\l$ in  \rf{1.6}    this
 matches the  structure of the  $1/N$  expansion of the  exact    $\mc N=4$ SYM  result in \rf{1.8}. 
One can also  use   similar considerations  to  predict the structure of   the string  theory  expansions  for other related observables 
%  starting  with  the  matrix model results
   \cite{Beccaria:2020ykg}. 

%This line of research has been extended in \cite{Beccaria:2020ykg} to other observables related to the BPS Wilson loops at the level of the matrix model formulation,  suggesting interesting 
%structures for the putative string perturbation expansion.
\def \cc {{\rm c}} \def \aa {{\rm a}}

It is important  to emphasize the universality of the  structure of the expansion in \rf{18}: it relies only on  the fact that one expands near  the AdS$_2$ minimal  surface 
embedded into the AdS$_3$ part of AdS$_n$ space  and  should   thus   be   valid  also for the corresponding 
 partition functions in the  ${\rm AdS}_{4}\times CP^{3}$  and ${\rm AdS}_{3}\times S^{3}\times T^4 $  superstring theories \cite{Giombi:2020mhz}.  It should   also apply to the orbifold   ${\rm AdS}_5\times (S^{5}/\mathbb Z_{2})$   theory:    
 orbifolding  the $S^5$  should not change the  above    argument     determining   the tension dependence from  the 
 way how  the AdS radius $L$  appears in   \rf{1.7}.\foot{In the case  of  $SU(N) \times ...\times SU(N)$  $\mc N=2$ 
 quiver  theory   which  is the $\mathbb  Z_{k}$  orbifold 
 of  the   $SU(k N)$  $\N=4$ SYM     and should be  dual to the    superstring on ${\rm AdS}_5\times (S^{5}/\mathbb Z_{k})$ 
 one has  for the AdS radius $\RR^4 = 4 \pi k N  \gs \a' $ and thus  instead of \rf{1.6}  we get 
 $\gs = \frac{g^{2}_{\rm YM}}{4\pi} = \frac{\lambda}{4\pi k N},\ \  T = { \RR^2 \ov 2\pi \a'}=\frac{\sqrt\lambda}{2\pi} 
$. 
   } 
   
   %AT10
   We thus conjecture  that  the  same form of the large $N$, strong coupling   expansion \rf{18}  or \rf{1.8}   should    also apply appear  in  
    the $\N=2$ orbifold theory case, i.e. 
    \be
\la{188}
\vev{\mc W}^{\rm orb} =  N\,e^{\sqrt\lambda}\sum_{p=0}^{\infty}
 \cc_p \frac{\lambda^{\frac{6p-3}{4}}}{N^{2p}}\Big[1+\mc O\Big(\frac{1}{\sqrt\lambda}\Big)\Big] \ , 
\ee
   where the coefficients  $\cc_p= { c_p \ov ( 8 \pi)^{p-1/2}}$ will  be    different from the ones  in \rf{1.8}.

In order to check the prediction \rf{18},\rf{188}   for  the  large $N$, strong-coupling expansion of $\vev{\mc W}^{\rm orb}$
we shall consider the  genus one term corresponding to the  leading $1/N^2$  correction
 to the planar    part   in \rf{1.1}.
%Turning to the orbifold theory that should be dual to the  type IIB superstring  on  ${\rm AdS}_5\times (S^{5}/\mathbb Z_{2})$
%the first non-trivial  question is   about  the  gauge coupling $\l$ or the  string tension $T$ dependence 
%of the first   $1/N^2$   correction to the planar SYM  term \rf{1.1}  in  $\vev{\mc W}^{\rm orb}$  in \rf{00}. 
Normalizing to  $\vev{\mc W}_0$    in \rf{1.1}  we have   in  both $\mc N=4$ SYM and $\mc N=2$ orbifold  cases
\be
\la{1.9}
\frac{\vev{\mc W}}{\vev{\mc W}_0} = 1+\frac{1}{N^{2}}\,q(\lambda)+\mc O\Big(\frac{1}{N^{4}}\Big), \qquad \qquad \vev{\mc W}_0=\frac{2N}{\sqrt\lambda}I_{1}(\sqrt\lambda)\ , 
\ee
where  the  form  of the  function $q(\l)$   will be our main  focus in what follows. 
In the    $SU(N)$  SYM case the expression for  $q(\l)$ follows from the expansion of  the exact Laguerre polynomial 
expression in \rf{1.8} ($I_n$ are modified Bessel functions of the first kind) 
\be
\la{1.10}
%\mc N=4:
\qquad q^{\rm SYM}(\lambda) = \frac{\lambda}{96}\Big[\frac{\sqrt\lambda\,I_{2}(\sqrt\lambda)}{I_{1}(\sqrt\lambda)}-12\Big] = \begin{cases}
-\frac{1}{8}\lambda+\frac{1}{384}\lambda^{2} +\mc O(\lambda^{3}), &\qquad  \lambda \ll 1, \\
\frac{1}{96}\lambda^{3/2}-\frac{9}{64}\lambda+\frac{1}{256}\lambda^{1/2}+\mc O(1), &\qquad  \lambda \gg 1 \ . 
\end{cases}
\ee
%that follows from the expansion, \cf (\ref{1.8}), \be
%\la{1.11}
%\vev{\mc W}^{\rm SYM} = N\,\frac{2}{\sqrt\lambda}I_{1}+\frac{1}{N}\frac{\lambda}{48}\Big(I_{2}-\frac{12}{\sqrt\lambda} I_{1}\Big)+\mc O(1/N^{3}).
%\ee
As  discussed   above, the leading  strong-coupling   behaviour of the genus one  correction   in $\vev{\mc W}^{\rm SYM}$
\be 
{ 1 \ov N^2}\,   q^{\rm SYM}(\lambda) \ \ \stackrel{\lambda\gg 1}{\sim}\ \   {  \lambda^{3/2} \ov N^2} \ {\sim} \ \  { \gs^2 \ov T} \ee 
is  consistent with  the universal form of the string theory expansion in  \rf{1.8}.
Then  according to \rf{888} the   same    should be true also   in the orbifold theory case, i.e.
%Since \rf{1.8}  follows from the  simple  world-sheet theory considerations (cf. \rf{1.7})  that should apply also to the 
%${\rm AdS}_5\times (S^{5}/\mathbb Z_{2})$   orbifold  string theory case  we expect   to find the same  scaling behavior also 
%in this  case, i.e. 
\be
\la{x1}
q^{\rm orb}(\lambda) \ \stackrel{\lambda\gg 1}{=}\  C\,\lambda^{3/2}+\mc O(\lambda)\  . 
\ee
 %showing consistence with the above heuristic argument. 
 where the  value of the coefficient $C$  may of course   be different from ${1\ov 96}$ in the SYM case in  \rf{1.10}.
Confirming the prediction  \rf{x1} 
starting from the  localization matrix model representation for $ \vev{\mc W}^{\rm orb}$  % in the $\mc N=2 $ gauge theory 
 will be one of the  aims   of the present  paper.

\subsection*{Summary of the results}

As we shall see below, the    matrix model representations for  the  orbifold $\mc N=2$  gauge theory   partition function $Z^{\rm orb}(\l; N)$  on $S^4 $  and  for    
$\vev{\mc W}^{\rm orb}$  imply  a remarkable relation  between 
$\Delta q(\lambda)$  %defined   as the difference of  $q^{\rm orb}(\lambda) $ and $   q^{\rm SYM}(\lambda)$   in \rf{1.9}, i.e. 
\be \la{x3}
\Delta q(\lambda) \equiv q^{\rm orb}(\lambda) - q^{\rm SYM}(\lambda)  \ , 
\ee
  and  the $N \to \infty$  limit of the  deviation of the orbifold   free energy $F^{\rm orb}= - \log Z^{\rm orb}$ 
%of the orbifold theory on $S^4$
   from its SYM counterpart 
%\be
%\la{x2}
%\Delta q(\lambda) = \lim_{N\to \infty}\frac{\lambda^{2}}{8}\frac{d}{d\lambda}\log\frac{Z^{\rm orb}(\lambda; N)}{[Z^{\rm SYM}(\lambda; N)]^{2}} 
%= - \lim_{N\to \infty }\frac{\lambda^{2}}{8}\frac{d}{d\lambda}\,\big[F^{\rm orb}(\lambda; N)-2\,F^{\rm SYM}(\lambda; N)\big]
%\ee
\ba
\la{x2}
\Delta q(\lambda) &= - \frac{\lambda^{2}}{8}\frac{d}{d\lambda}\,\Delta F(\lambda)\ , \\
\Delta F(\lambda) &\equiv    \lim_{N\to \infty }\big[F^{\rm orb}(\lambda; N)-2\,F^{\rm SYM}(\lambda; N)\big] = -\lim_{N\to \infty}\log\frac{Z^{\rm orb}(\lambda; N)}{[Z^{\rm SYM}(\lambda; N)]^{2}}  \ . \la{1.16}
\ea
%where $F(\lambda; N)$ denotes the free energy $F = -\log Z$. 
The leading $\mc O(N^{2})$  term  in  $F^{\rm orb}-2\,F^{\rm SYM}$  cancels 
due to the planar equivalence  between the $SU(N) \times SU(N)$ orbifold  theory  and  the  two decoupled copies
 of the  $SU(N)$ $\mc N=4$ SYM theory.\footnote{More precisely, 
for the Wilson loops  \rf{00}
 the  planar equivalence means that  the  normalized expectation   value of  an operator  in one of the two $SU(N)$  factors of the quiver  is 
the  same as in  the  $SU(N)$ $\mc N=4$ SYM theory. 
In   general, the planar correlators of operators   from $\mathbb Z_2$ symmetric (i.e. ``untwisted'')  
 sector should match  between 
the orbifold and the  parent $SU(2N)$ SYM theory. 
For  ``extensive'' quantities like the  free energy (or conformal anomalies,  correlators of total stress tensor, {\em etc.})
the  $N \to \infty$ results in the $SU(N)\times SU(N)$  orbifold  theory  should
match those  of the 
 two copies of the  $SU(N)$  $\mc N=4$ SYM. 
}
%The remainder is finite and non-trivial when $N\to\infty$.
%Eq.~(\ref{x2}) follows  from the structure of the localization matrix model 
%(so effectively from $\mc N=2$  supersymmetry). 
\def \ff {{\rm f}}
Using \rf{x2} the  expected strong-coupling   behaviour  \rf{x1}  of $q(\l)$  
translates into the following  scaling    for the difference of free energies in \rf{1.16}  ($c_1=-16 C$)
\be
\Delta F(\lambda) \stackrel{\lambda\gg 1}{=} \  c_1\,   \lambda^{1/2} + ... \ . \la{1.17}
\ee
In view of \rf{1.16} this  leads to a prediction  about the strong coupling asymptotics of  the leading $1/N^2$ correction 
to the planar  $F^{\rm orb}(\lambda; \infty)=2\,F^{\rm SYM}(\lambda; \infty)$  part of  the  free energy of the orbifold theory.

%%%%%%%%%%%%%%%%%%%%%%%
Let us first  recall that 
 the  free  energy of $SU(N)$  $\mc N=4$  SYM on $S^4$  should not  be  renormalized, i.e. should be  given exactly 
by the  familiar  one-loop expression % cutoff-dependent  (one-loop)  result, 
$F^{\rm SYM}(\lambda; N; \Lambda) =  4\,\aa\, \big[\log ({\rr\Lambda}) + f_0 \big]$. Here  $\aa$  is the conformal anomaly 
coefficient, $\aa=\frac{1}{4} (N^{2}-1)$,  $\rr$ is the radius of $S^4$, $\Lambda$  is  a 4d  UV cutoff 
and $f_0$  is a constant.  Since the free  energy is UV divergent, its finite part is not universal depending on a particular  
regularization scheme. The localization procedure \cite{Pestun:2007rz} representing  the  
free energy  in terms of a  finite   matrix model integral with a  simple  $\l$-independent measure  implicitly assumed  a special 
regularization in which the  renormalized $SU(N)$   SYM free energy is given by 
\be \la{118}
F^{\rm SYM}(\lambda; N) = - 2  \aa \log { \lambda} =- \ha (N^2-1) \log {  \l}     \  \ee
as this expression  follows simply from the Gaussian matrix model integral  \cite{Russo:2012ay} 
(we drop  an additive numerical constant).  

From the dual string theory point of view, the gauge theory  free energy  is expected to be  given by the 
\adss   string  partition   
%v4
function  or   (at the tree level)  by the IIB  string effective action
$S=S_0 + S_1 + ...= { 1 \ov (2 \pi)^7 \gs^2 \a'^4} \int d^{10} x \sqrt G\big[ ( R +...)  +   \a'^3  R^4 + ...\big]   + \mc O(\gs^0)+...$. 
Evaluated   on \adss (using \rf{1.6}  and $R+...={-8 L^{-2}}$)    the leading supergravity term    
 here   is  $S_0={1 \ov \pi^2}  N^2  V_{\rm AdS_5}$ 
where $V_{\rm AdS_5}$  is the (logarithmically)  IR divergent  volume of  unit-radius  ${\rm AdS_5}$.
Subtracting the  IR divergence   in   $V_{\rm AdS_5}$     using a particular  AdS/CFT motivated prescription 
% (with an  effective  length or inverse mass cutoff  rescaled  by $\sqrt \lambda$) 
 gives $V_{\rm AdS_5} \to  \bar V_{\rm AdS_5}= -\pi^2  \log \sqrt \lambda $ 
 and thus one  reproduces    \cite{Russo:2012ay}  the $N^2 \log \l$ term  in \rf{118}.\foot{One way to   understand 
the origin  of the $\log \sqrt \lambda$   term is as follows. On AdS side the IR cutoff   is  measured in units  of  AdS radius $L$.  On the   gauge theory side viewed as originating from the  flat-space open string theory  the natural UV cutoff is  inverse of the 
 string length $\sqrt { \a'}$.
Thus the two cutoffs are related  by the ratio $ {L\ov \sqrt { \a'}}= \l^{1/4}$. 
}

 The $-1 $   shift of $N^2$  in \rf{118}    should come from the  1-loop  superstring correction (again proportional to $V_{\rm AdS_5}$): this  
should     follow the same pattern as 
 found for the $\N=4$ SYM  conformal anomaly and  $S^3$  Casimir   energy  in \ci{Beccaria:2014xda}
 (with only loops of short  supergravity supermultiplets  contributing). 
 Other   string $\a'^n$  tree level (e.g. $\a'^3 R^4 $, cf. \ci{Gubser:1998nz})    and  
 string loop corrections should   vanish on  maximally supersymmetric  \adss  background.

% The  leading $N^2$  term in this localization  matrix model expression matches   the on-shell value of the  leading (supergravity) term in the string theory  effective action in 
% \adss under a natural   choice of the  IR  cutoff in the  ${\rm AdS}_5$  volume \cite{Russo:2012ay}.
 %\foot{Starting from the matrix model representation ref. \cite{Russo:2012ay} found also  the strong coupling limit of the leading  large $N$ term  in the  free energy  of the  superconformal $\mc N = 2$ $SU(N)$   gauge theory  theory with $N_{f} = 2N$ fundamental  hypermultiplets.}

 Let us now turn to the orbifold theory   that should be dual  to the   superstring on ${\rm AdS}_5\times (S^{5}/\mathbb Z_{2})$. 
   Combining \rf{1.16},\rf{1.17} and \rf{118}   we get the following prediction 
 \be \la{119}
F^{\rm orb}(\lambda; N) \stackrel{\lambda\gg 1}{=} - N^2 \log { \l}   +  \big[ c_1\,    \lambda^{1/2} +  \mc O(\log \l)   \big] 
+ \mc O \Big({1\ov N^2}\Big)    \ . \ee
The    leading $N^2$   term here  is  implied by  the  planar equivalence  to 
  the $SU(2N)$   SYM   and should follow  again  from the leading type IIB supergravity  term  evaluated on 
  ${\rm AdS}_5\times (S^{5}/\mathbb Z_{2})$.\foot{To recall,   the orbifold projection of   $SU(2N)$    SYM giving the $SU(N) \times SU(N)$  theory with two bi-fundamental hypermultiplets reduces the number of degrees of freedom and thus also the leading
 %v4
  large $N$  term in the conformal anomaly
coefficient from  $\aa={ 1\ov 4}  (2N)^2  $ to  $ 2 \times  { 1\ov 4}  N^2$
 which is twice the anomaly of a single copy of  $SU(N)$ SYM theory at large $N$. 
 %v4
 The exact  expressions for the conformal anomaly   coefficients of the $SU(N) \times SU(N)$     quiver theory are 
 $\aa= \ha N^2 - {5\ov 12}$ and $\cc =  \ha N^2 - {1\ov 3}$.
On the supergravity side, replacing  the $N^2$ coefficient  in the above discussion  by $ (2N)^2$   and 
noting that the volume  of $ (S^{5}/\mathbb Z_{2})$    is   half   of  the volume of $S^5$    we end up with $2 N^2$  as an overall coefficient.
    }
The planar equivalence    also   means  that   like in \rf{118} 
 this  leading $N^2$ term should not get    string tree level $\a'$-corrections, i.e.  they should still vanish when evaluated on  
${\rm AdS}_5\times (S^{5}/\mathbb Z_{2})$.

One  may attempt  %Remarkably, it  may be possible 
 to give an independent string theory explanation of the
subleading $\l^{1/2}$  term in \rf{119}  without using the connection \rf{x2}  to the Wilson loop.
 The string  one-loop  (genus one or order $\gs^0\sim N^0$) 
 type IIB    effective action  is known to start with 
    $S_1\sim  {1\ov \a'} \int d^{10} x \sqrt G\,   R^4 +...$   \cite{Green:1982sw,Gross:1986iv,Sakai:1986bi}.
    If  we conjecture that  when evaluated on   the orbifold  ${\rm AdS}_5\times (S^{5}/\mathbb Z_{2})$ 
    it is  no longer zero
    % (some cancellations may  not happen due to reduced supersymmetry), 
     then  on dimensional grounds it should scale as  
    $S_1 \sim {L^2\ov \a'} \sim \sqrt \l $,    reproducing the  subleading  term  in \rf{119}. If a non-zero contribution  comes just  due to the  curvature singularity  then it may not be  proportional to the AdS$_5$  volume so there will be no extra $\log  { \lambda}$ factor.
   %   \foot{One potential   problem with this argument 
   %  is that  one may expect also a factor of  $V_{\rm AdS_5}$   which,  if defined in the same scheme    as  in the leading order contribution,  would 
%     introduce an extra $\log \l$ factor.  Another  
  The remaining puzzle is  why    the one-loop $R^4$ term may contribute to $F^{\rm orb}$    while the   tree-level   one should not, 
  even though the two invariants have the same structure in type IIB   string theory \ci{Kiritsis:1997em}
  %v3
  (cf. the case of  compactification of 6d orbifolds   \ci{Antoniadis:2002tr}).

 %     question    to explain  is  why the tree level $\a'^3 R^4$ term does not  contribute on 
  %  ${\rm AdS}_5\times (S^{5}/\mathbb Z_{2})$  (it would lead to an extra $\l^{-3/2}$ correction to $N^2$ term in \rf{119}) 
   %         while the similar  one-loop  term  does.}
\iffa 
\begin{verbatim}
ume scales as L^10 so S_1 ~  
  \end{verbatim}
\cite{Green:1982sw,Gross:1986iv,Sakai:1986bi,Metsaev:1987ju,Gubser:1998nz,Beccaria:2014xda}
\fi 

\

%%%%%%%%%%%%%%%%%%%%%%%%%%%%%%%%%%%%%%%%%%%%%%%
Starting   with the matrix model representation  for $\vev{\mc W}^{\rm orb}$  we 
 shall first study  the structure of the function $q^{\rm orb}(\lambda)$   in \rf{1.9}  or $\Delta q(\lambda)$ in \rf{x3} 
    at     weak coupling.  While 
    the small $\l$   expansion of $q^{\rm SYM}(\lambda)$  in \rf{1.10}   has only rational coefficients, 
     the coefficients  in the expansion  of $\Delta q(\lambda)$ 
    in powers of $\l$ are transcendental  -- proportional to  the  values  $\zeta_{n} \equiv  \zeta(n)$
    of the Riemann $\zeta$-function
 %     finding that
%To investigate the strong coupling regime in (\ref{x1}) one needs to collect as much as possible information 
% about the function $q(\lambda)$. In weak coupling perturbation theory one finds the expansion 
 \ba
% \la{x3}
%q^{\rm orb}(\lambda) &= q^{\rm SYM}(\lambda) + \Delta q(\lambda),   \\
 \tfrac{1}{8\pi^{2}}\, \Delta q(\lambda) &= -\tfrac{3}{4}\zeta_{3}\, \Big(\frac{\lambda}{8\pi^{2}}\Big)^{3}
 +\tfrac{45}{8}\zeta_{5}\, \Big(\frac{\lambda}{8\pi^{2}}\Big)^{4}
+\Big[\tfrac{9}{2}\zeta_{3}^{2}-\tfrac{315}{8}\zeta_{7}\Big]\, \Big(\frac{\lambda}{8\pi^{2}}\Big)^{5}+\mc O ( \l^6),  \la{1.14}
\ea  
%Here   $\zeta_{n} = \zeta(n)$ are the Riemann $\zeta$-function values. 
Thus  $\zeta_n$  may be formally used    to  parametrise   the  deviation of the orbifold  theory result  
from  the $\mc N=4$ SYM one. 

To find the strong-coupling expansion of $q^{\rm orb}(\lambda) $    requires a resummation of the 
%This works well at weak coupling, but a resummation is necessary to capture 
%the strong coupling limit.
 weak coupling expansion.
We shall  study resummations of particular subclasses   of terms  proportional to monomials built out of $\zeta_n$.
While this will not be enough to determine the correct strong coupling asymptotics of $q^{\rm orb}(\lambda)$,   this  may still 
help to  shed light on  the general structure of this function.\footnote{A similar approach was  applied 
  \cite{Beccaria:2020hgy} to 
 $SU(N)$ superconformal $\mc N=2$ theories admitting a large $N$ limit. 
Also, the use of    sufficiently  many terms in the 
  perturbative   series  as a guide  towards 
some  non-perturbative features  like singularities was  emphasised in \cite{Fiol:2020ojn,Fiol:2020bhf}.}

Exploiting the relation (\ref{x2}) we shall compute $\Delta q(\lambda)$ up to order $\mc O(\lambda^{20})$ and also determine the resummation of all terms  with  the following  types of  coefficients  involving particular $\zeta_n$  and their powers
%with the following structure 
\be
\text{I:}\ \zeta_{2n+1}, \quad 
\text{II:}\ \zeta_{3}\,\zeta_{2n+1}, \quad
\text{III:}\ \zeta_{3}^{p},\quad 
\text{IV:}\ \zeta_{3}^{p}\zeta_{5}^{q}, \quad
\text{V:}\ \zeta_{3}^{p}\zeta_{5}^{q}\zeta_{7}^{r}\ , 
\ee
%Denoting  by $q^{\rm orb}_k$ ($k=\text I, ..., \text V$)    the  corresponding  subsets of terms  in $q^{\rm orb}$
i.e.  $q^{\rm orb}_{\rm I} = \sum  c_{n, m} \zeta_{2n+1}  \l^m$, etc.
We find that   they have the following behaviour at   strong coupling  % ($\l \gg 1$) 
\ba
\la{x4}
& q^{\rm orb}_{\rm  I}(\lambda) \stackrel{\lambda\gg 1}{=} -\tfrac{3}{128\,\pi^{2}}\,\lambda^{2}+\mc O(\lambda^{3/2}), \qquad\qquad 
q^{\rm orb}_{\rm  II}(\lambda) \stackrel{\lambda\gg 1}{=} -\tfrac{9\,\zeta_{3}^{2}}{4096\,\pi^{8}}\,\lambda^{5}+\mc O(\lambda^{4}),\\
%\ee
%while in the cases III, IV, V we have the $K=1,2,3$ instances of the general formula
%\be
\la{x5}
& q^{\rm orb}_{\rm III,IV,V }(\lambda) \stackrel{\lambda\gg 1}{=} -\tfrac{(k+1)(2k-1)}{16}\, \lambda + \mc O(1)\ , \ \ \ \ \ \ \ \ \ \ \ 
k= {\rm 1, 2, 3} \ . 
\ea
% valid when taking into account any monomial in the first $K$ $\zeta$ numbers.
The difference  in these asymptotics  implies  that  to find the  correct strong coupling behaviour of  the full 
$q^{\rm orb}$ 
one  needs   first  to   sum  together    different subsets of terms   and only then expand the total  at large $\l$.

%The difference between the asymptotics in cases I and II already shows that it is far from trivial to 
%identify the relevant contributions in the weak coupling series. For instance, the $\sim \lambda^{5}$ behaviour in case II
%is hardly achievable on string side and shows that cancellations are at work at strong coupling.
%Even worse, from (\ref{x5}) we see that, although taking $K\to\infty$ will converge to the exact result at finite $\lambda$ (by construction since we are
%keeping all terms), it is not possible to exchange it with the limit $\lambda\to\infty$. 

\

Lacking an analytic  method   to compute  $q^{\rm orb}(\l)$ %\Delta q(\lambda)$ 
at  strong coupling
we performed extensive numerical simulations of the   $SU(N) \times SU(N)$ 
orbifold matrix model to measure  it. %$q^{\rm orb}(\lambda)$. 
This required an extrapolation to large $N$  for finite 
 $\lambda$, followed by 
an analysis of  large $\lambda$  region. %At the $\mathbb Z_{2}$ symmetric point, 
We confirmed that the deviation from the $\mc N=4$ SYM 
 case  starts only at the non-planar level.
The numerical data agrees with  the Pad\'e-Borel resummation of the weak-coupling expansion up to moderate $\lambda \sim 50$. At larger values of  the coupling $\l$  we found that the  data  
%, we  provide support to the conclusion that $q(\lambda)$ 
 is compatible with the following  asymptotics 
\ba
\la{x7}
&\qquad \qquad q^{\rm orb}(\lambda)   \ \stackrel{\lambda\gg 1}{=} C \,  \lambda^{\eta} \Big[ 1 + { a_1\lambda^{-1/2} } + ...\Big] \ , 
%\  C_{0}\, \lambda^{\eta}+C_{1}\,\lambda^{\eta-\frac{1}{2}}+\cdots,
\\
&\la{xx7}
\eta = 1.49(2) \ , \ \qquad \ \ \  \ \ C\simeq - 0.0049(5)\ , \ \ \ \ \ \ \ \ \  a_{1}\simeq    15.5(5) \ .
%-7\cdot 10^{-2} \ , 
\ea
%with  coefficients $C_{0}\simeq -4.5\cdot 10^{-3}$, $C_{1}\simeq -7\cdot 10^{-2}$,
% and an exponent which we estimate to be  $\eta = 1.49(2)$, thus close to $\frac{3}{2}$, \ie 
%AT
The power  of the   leading asymptotics  $\eta\approx 1.5$   is thus   consistent  with  the   string theory prediction (\ref{x1}).\footnote{
If $\eta$ is set to be exactly  $3/2$, then the best fit value of the coefficient $C$ slightly changes to $-4.7\cdot 10^{-3}$.}
It is interesting to notice that the  values of  the coefficients  $C$ and   $C  a_1$  are  very  close to the values 
of the  corresponding coefficients in \rf{1.10} in the SYM case  up to a factor of $-\ha$ and $+ \ha$ respectively. 
This suggests a conjecture  that the exact   form of the strong coupling expansion of $q^{\rm orb}(\lambda) $ is given by 
\be \la{xxx7}
q^{\rm orb}(\lambda)   \ \stackrel{\lambda\gg 1}{=}  -  \tfrac{1}{192}\lambda^{3/2} - \tfrac{9}{128}\lambda
%{1\ov 2} \big(-  \frac{1}{96}\lambda^{3/2} - \frac{9}{64}\lambda\big) 
+\mc O(\lambda^{1/2}), 
\ee
It remains to be  seen if  one can  prove this analytically. 

\

%AT10
One may wonder if 
 the   coefficient of the  leading $\l^{3/2}\ov N^2$  correction  in $\vev{\mc W}^{\rm orb}$ may be 
 found, as  in the  $\N=4$ SYM case  \ci{Drukker:2005kx},   also 
   by  considering the  circular   Wilson loop in $k$-symmetric  representation
   which should  be described  (for  $k \gg 1$  and $k \sqrt{\l}\ov N$=fixed) by a classical D3-brane solution. % in AdS$_5$.
 %and then interpolating to smaller values of $k$. 
 This  does  not   seem  possible   as the D3-brane solution  of 
 \ci{Drukker:2005kx} is  restricted to AdS$_5$   and  thus  the  $ k^3 \l^{3/2}\ov N^2$  term  in its action   should have the same 
  coefficient   as in the $\N=4$ SYM case, in contradiction with \rf{xx7},\rf{xxx7}. 
 In  fact, the D3-brane solution  of \ci{Drukker:2005kx}  should   be related  not to the   $SU(N)$  Wilson loop \rf{00} 
 of the $SU(N) \times  SU(N)$ orbifold theory but  to the    orbifold projection of the original  circular Wilson loop 
 in  the $SU(2N)$   SYM theory. The projection of the  latter    is represented by 
   the correlator $\vev{\mc W_1 \mc W_2}$  where $\mc W_{1,2}$   in \rf{0} 
 correspond to the two $SU(N)$ factors  of  the orbifold theory.  
 %The correlator  $\vev{\mc W_1 \mc W_2}$    factorizes 
 %in the planar   limit  and one might  first   think  that  in view of the above discussion 
%  the   subleading  $\l^{3/2}\ov N^2$  term in it    should   then be the same as in the SYM case. 
  %However,  by 
  Starting with  the   $SU(2N)$  Wilson loop in $k$-symmetric representation   one  is to split it into  the  sum of 
  products of the two  % ($m$-symmetric and $(n-m)$-symmetric)  
  $SU(N)$  representations. Then  the  D3-brane  description  may  apply only 
  to a  special   combination  of the $\vev{\mc W_1 \mc W_2}$  correlators where  $\mc W_1$ and $\mc W_2$ 
  are taken in the particular   representations  %(including trivial one) 
  of the $SU(N)$ appearing in the product.\foot{We   thank N. Drukker for this suggestion.}
 %   We leave the   study  of such correlators  for  the future. 
 % Interpolating to smaller  values of $k$ and 
 Assuming that $k$-symmetric representation   may be replaced 
 by the $k$-fundamental one (corresponding to multiply wrapped circle, cf. \cite{Hartnoll:2006is,Yamaguchi:2007ps}) 
 one would expect to get  the sum  of   the correlators   
 \be \la{XXX}
 \sum_{m=0}^k  \binom{k}{ m}    \langle \mc W^{(m)}_1 \mc W^{(k-m)}_2\rangle\ee
 where $\mc W^{(m)}_\aa$   is the $SU(N)$   Wilson loop in $m$-fundamental representation. 
% where  each factor  has  different  number of wrappings ($k_1, k_2=0, ...,k$)
 Rescaling the  fields in \rf{00}  (or the corresponding matrices in the matrix model representation  as in \ci{Drukker:2000rr})
 one would then   end up with the   sum of the correlators  $\vev{\mc W_1 \mc W_2}$  of the two fundamental  $SU(N)$ Wilson loops 
 in the  $SU(N) \times SU(N)$  quiver   theory with the two 't   Hooft  couplings
 $\l_1 = m^2 \l,\  \l_2 = (m-k)^2 \l$.  The resulting expression  should   simplify in the large $k$ limit  
 %The study of  such correlators  is an interesting problem for the future. 
 and one  expects it to be  dominated by the  ``diagonal''  term   ($m=k/2$)    with $\mc W_\aa$ in the same  representation. 
 In section 6  we shall present     numerical  data    indicating      that the $1/N^2$ term in this  correlator (with both Wilson loops taken in the fundamental  representation of $SU(N)$)  has  a  similar 
 strong-coupling behaviour to  $ {1\ov 96} \l^{3/2}$  found  in the SYM case in \rf{1.10}.

\

Below  we    also computed   numerically   the  individual  % the  matrix model representation for  the 
  $SU(N)$ Wilson loop   \rf{0}  expectation values $\vev{\mc W_{1}}, \vev{\mc W_{2}}$ 
   in the  $SU(N) \times SU(N) $  quiver  with unequal  couplings $\l_1,\l_2$
   starting with the localization matrix model representation. 
 %Considering a particular value  of the  ratio  of the couplings 
% $\lambda_{2}/\lambda_{1}=3$ and 
%Guided by   the  strong-coupling  results of \cite{Zarembo:2020tpf} 
  Guided by the discussion in     \cite{Zarembo:2020tpf}  here   we considered  the following  analog  of  the ratio in (\ref{1.9})\foot{$\vev{\mc W_{2}}$  is found  by interchanging  $\lambda_{1}$ and $\lambda_{2}$
or $\theta \to 2 \pi -\theta$.}  
\be\la{125}
\frac{\vev{\mc W_{1}}}{w(\theta)\,\vev{\mc W}_0} = p(\lambda; \theta)+\frac{1}{N^{2}}\,q(\lambda; \theta)+\mc O\Big( {1\ov N^4}\Big) \ , 
\ee
where
\be\la{126}
\lambda =  \frac{2\lambda_{1}\lambda_{2}}{\lambda_{1}+\lambda_{2}}, \qquad \ \ \ \ \theta = 2\pi\,\frac{\lambda_{1}}{\lambda_{1}+\lambda_{2}}, \qquad \ \ \ \ w(\theta) = 
\frac{1-\frac{\theta}{2}\cot\frac{\theta}{2}}{\sin^{2}\frac{\theta}{2}} \ , 
\ee
%so that  according to the strong coupling result of   \cite{Zarembo:2020tpf}   we should have 
%   $ p(\lambda; \theta) \to 1 $   at $\l\to \infty$. 
 The strong coupling result  of \cite{Zarembo:2020tpf} 
then implies  that   $p(\lambda; \theta)\big|_{\l \to \infty} \to 1$  %at large $\lambda$,
 for any $\theta\neq 0, 2\pi$. 
We confirmed this prediction numerically by   considering a   particular value  of the  ratio  of the two couplings 
 $\lambda_{2}/\lambda_{1}=3$ (i.e. $\theta = {\pi \ov 2}$) 
%MB1 
 and measuring both Wilson loops $\vev{\mc W_{1}}$ and $\vev{\mc W_{2}}$, thus 
 effectively probing also  the value $\theta=\frac{3\pi}{2}$.
%, at least within the achieved accuracy. 
%More generally,    
%MB1
We found  that   the  strong-coupling   expansion of the function $p$ 
 has the form  $ p(\lambda; \theta) =1 + h(\theta)/\sqrt\lambda + ...$ 
where $h$  has  a  non-trivial dependence on $\theta$. % (so it  is different in the case of $\vev{\mc W_{2}}$). 
%depends on Corrections are $\mc O(1/\sqrt\lambda)$ and depend on $\theta$, \ie are different for the two Wilson loops.
The  numerical data   for the function $q(\lambda; \theta)$ turns out to be  compatible with the strong-coupling 
asymptotics  in  (\ref{x7})
with the  exponent  $\eta$  being  again close to $3/2$ % (within errors)
 independently of $\theta$, i.e. for both Wilson loops.

\

The rest of this paper is organized as follows.  In section 2   we shall   present  the matrix model integral  representation for the Wilson loop  expectation value in the quiver theory  which will be our starting point.  
In section 3 we shall  consider  the  weak coupling expansion of the leading non-planar term   the  case of the 
orbifold  theory organising  it  in terms of monomials of transcendental $\zeta_n$  factors. We shall  also derive the relation between 
the function $q(\l)$ and the free energy. In section 4 we shall perform a resummation of some subsets of terms  and  then expand them at strong coupling
finding non-universal  behaviour.  In section 5 we shall consider  the  weak-coupling expansion in the case of  non-symmetric 
 quiver  theory. Finally, in section 6 we shall present the results of the numerical  computation of the  matrix model integrals. 
 Appendices A and B  will contain some technical details of the computations in sections 3 and 4. In Appendix C we shall briefly discuss   similar weak-coupling analysis  of the Wilson loop in   $SU(N)$  ``orientifold''   $\mc N =2$ superconformal  theory.

\paragraph{Note added in v4:} 
The coefficient $c_1$ in (\ref{1.17})   was  recently computed
exactly in \ci{Beccaria:2022ypy}  with the result $c_1 = {1\ov 4}$
(also, coefficients of several subleading terms in \rf{119}  were also found). 
 Since according to \rf{x2} 
 the   coefficient  $c_1$  is related to the coefficient
  of $\l^{3/2}$ term in $q$ or $C$ in   (\ref{x7})  as  $C=-{1\ov 16} c_1$ 
  this implies that  $C=-{1\ov 32}$ (invalidating the conjecture in \rf{xxx7}).
   This value %(-0.03125)
is substantially larger than the  numerical estimate for $C$   in (\ref{xx7}). 
This shows that the explored range of 
$\lambda$ values was  too narrow to reach the asymptotic regime $\lambda \gg 1$ where
the leading $\l^{1/2}$ term dominates. 
This is consistent with the  %fact that in \red{[new paper]}
presence of  the  subleading $\log \l$  correction in $\Delta F$  in \rf{1.17} (also established in 
\ci{Beccaria:2022ypy})  that slows  down the  convergence.

%%%%%%%%%%%%%%%%%%%%%%%%%%%%%%%%%%%%%%%%%%%%%%%
\section{Matrix model representation }

Our   starting point will be the localization  matrix model representation for the $S^4$ partition function  and the 
expectation   values of the  circular Wilson loops \rf{0}   in the   $SU(N) \times SU(N) $  $\mc N=2$    superconformal quiver  theory
 \ci{Pestun:2007rz} (see also \ci{Rey:2010ry,Passerini:2011fe,Zarembo:2020tpf}).
 The partition  function  may be   written  as the   integral over two sets of eigenvalues     ($\aa=1,2; \ i=1, ..., N$)
\be
\la{2.1}
Z = \int \prod_{\aa=1}^{2}\Big[\prod_{i=1}^{N}da_{\aa i}\ \delta(\sum_{i} a_{\aa i})%\delta(\sum_{i} a_{2i})
\, \prod^N_{i<j}(a_{\aai}-a_{\aaj})^{2}
\ e^{-\frac{8\pi^{2}N}{\lambda_{\aa}}\sum_{i}a_{\aa i}^{2}}
\Big] \, f[a_{1 }, a_{2 }]\, 
%\frac{\prod_{a}\prod_{i<j}(a_{ai}-a_{aj})^{2}\,H^{2}(a_{ai}-a_{aj})}{\prod_{ij}H^{2}(a_{1i}-a_{2j})}
\ ,
\ee
where the $\delta$-functions reflect the  fact that we are considering the $SU(N)$ case (they may  be ignored  in strict planar limit)   and\footnote{\la{f1}
 $H(x)$  has   the following representation in terms of the Barnes function  $G(x)$
\be\notag 
\log H(x) = \log \big[G(1+i x)G(1-ix)\big]-(1+\gamma_{\rm E})\,x^{2}.
\ee
%where $G$ is the Barnes function. 
The partition function is invariant under 
$H(x)\to H(x)\, e^{Cx^{2}}$  \cite{Passerini:2011fe}. 
%where it is explained in terms of UV finiteness.
%AT
Note also    that we ignored the instanton factor  \ci{Pestun:2007rz}   in the integrand as   we will be interested in 
perturbative $1/N$  expansion  (see \ci{Passerini:2011fe}).
} 
%\footnote{Notice that we need explicit delta-function to enforce the $SU(N)$ constraint. This gives corrections that drops at large $N$ but  modify the subleading terms of the $1/N$ expansion.}
\be
\la{2.2}
 f[a_{1 },a_{2}] = \frac{\prod_{\aa}\prod_{i < j} \,H^{2}(a_{\aai}-a_{\aa j})}{\prod_{i,j}H^{2}(a_{1i}-a_{2j})}\ \ , \qquad \qquad 
H(x) \equiv  \prod_{n=1}^{\infty}\Big(1+\frac{x^{2}}{n^{2}}\Big)^{n}\, e^{-\frac{x^{2}}{n}} \ .
\ee
Below we shall 
   use  $\vev{\cdots}_0$  to denote the  (normalized)   expectation value in 
   the  matrix model   with the   Gaussian measure 
% $e^{-\tr (A_{1}^{2}+A_{2}^{2})}$ (or $e^{-\tr A_{1}^{2}}$  after we integrate out $A_2$),  
 so that 
\rf{2.1}    may be written as 
\ba\la{211}
&\qquad \qquad Z=  Z_0  \vev{f }_0 \  , \qquad \qquad     Z_0=  Z^{\rm SYM}(\l_1;N)\,  Z^{\rm SYM}(\l_2;N)\ ,  \\
&  Z_0 =  \int \prod_{\aa=1}^{2}\Big[\prod_{i=1}^{N}da_{\aa i}\ \delta(\sum_{i} a_{\aa i})%\delta(\sum_{i} a_{2i})
\, \prod^N_{i<j}(a_{ai}-a_{aj})^{2}
\ e^{-\frac{8\pi^{2}N}{\lambda_{\aa}}\sum_{i}a_{\aa i}^{2}} 
\Big] \,
%\frac{\prod_{a}\prod_{i<j}(a_{ai}-a_{aj})^{2}\,H^{2}(a_{ai}-a_{aj})}{\prod_{ij}H^{2}(a_{1i}-a_{2j})}
  \ . \la{212} 
\ea
Here    $Z^{\rm SYM}(\l;N)$  is the  $SU(N)$ SYM partition function on $S^4$.

The expectation values of the two Wilson loops \rf{0}   are given by 
\be
\la{2.3}
\vev{\mc W_{\aa}}=\langle\, {\sum_{i=1}^{N} e^{2\pi a_{\aai}}}\,  \rangle = \frac{ \langle\, {f  \, \tr\,  e^{2\pi a_{\aa}}}\,  \rangle_0}{ \vev{f}_0}   \  , 
\ee
where $\langle ... \rangle$   is  given  by the same integral as in \rf{2.1}   and  is normalized  so that   $\langle  1  \rangle=1.$
We  use the notation $a_\aa$   for the  diagonal matrix $a_\aa= {\rm diag}(a_{\aa 1}, ..., a_{\aa N})$. 
The  two  expectation values \rf{2.3} 
are equal (cf. \rf{00}) at the orbifold point  $\lambda_{1}=\lambda_{2}=\lambda$. 
%\be\vev{\mc W_{1}}=\vev{\mc W_{2}}\stackrel{\rm def}{=} \vev{\mc W}. \ee

For   large $N$  one may study the saddle  points of the ``effective action'' in \rf{2.1} 
\be\la{25}
S[a_1,a_2] = N  \sum^2_{\aa=1}\frac{8\pi^{2}}{\lambda_{\aa}}\sum_{i}a_{\aai}^{2}- \log f[a_1,a_2] \ . 
%\sum_{a}\sum_{i<j}\Big[\log(a_{ai}-a_{aj})^{2}+\log H^{2}(a_{ai}-a_{aj})\Big]+\sum_{ij}\log H^{2}(a_{1i}-a_{2j}).
\ee
Differentiating over $a_{\aai}$   and introducing the densities   
\be
\la{2.5}
\rho_{\aa}(x) = % \vev
\langle\,  {\frac{1}{N}\sum_{i=1}^{N}\delta(x-a_{\aai})}\, \rangle \ ,
\ee
one finds the following saddle point equations  \ci{Rey:2010ry,Passerini:2011fe}
\ba
\la{2.6}
& \int_{-\mu_{1}}^{\mu_{1}}dy\,\rho_{1}(y)\,\Big(\frac{1}{x-y}-K(x-y)\Big)+\int_{-\mu_{2}}^{\mu_{2}}dy\,\rho_{2}(y)\,K(x-y) = \frac{8\pi^{2}}{\lambda_{1}}\,x, \\
& \int_{-\mu_{2}}^{\mu_{2}}dy\,\rho_{2}(y)\,\Big(\frac{1}{x-y}-K(x-y)\Big)+\int_{-\mu_{1}}^{\mu_{1}}dy\,\rho_{1}(y)\,K(x-y) = \frac{8\pi^{2}}{\lambda_{2}}\,x,\la{26} \\
%\ea where \be
\la{2.7}
&K(x) \equiv  -\frac{H'(x)}{H(x)} = x\,\big[\psi(1+ix)+\psi(1-ix)+2\gamma_{\rm E}\big] = -2\sum_{n=1}^{\infty}(-1)^{n}\,\zeta_{2n+1}\,x^{2n+1}\ .   
\ea
%which has  the following  small $x$ expansion 
%\be
%\la{2.8}
%K(x) = -2\sum_{n=1}^{\infty}(-1)^{n}\,\zeta_{2n+1}\,x^{2n+1}.
%\ee
The large $N$ equivalence  of the orbifold theory with the $\mc N=4$ SYM   follows \ci{Rey:2010ry,Zarembo:2020tpf} 
 from the fact  that 
for $\lambda_{1}=\lambda_{2}=\lambda$ the equations (\ref{2.6}),\rf{26}  admit the symmetric Ansatz $\rho_{1}=\rho_{2}=\rho$, $\mu_{1}=\mu_{2}=\mu$ and reduce to the saddle point equation 
for the Gaussian matrix model  corresponding to   the   $\mc N=4$ SYM  case for which 
\be
\rho(x) = \frac{2}{\pi\mu^{2}}\sqrt{\mu^{2}-x^{2}},\qquad\qquad  \mu = \frac{\sqrt\lambda}{2\pi}\ .
\ee
The solution of the two integral equations (\ref{2.6}),\rf{26}  in the large $\lambda$   limit was  studied  in \cite{Rey:2010ry},
showing   that $\vev{\mc W_{a}}\sim e^{\sqrt\lambda}, \ {\lambda} = \frac{2 \l_1 \l_2 }{\lambda_{1} +\lambda_{2}}$, 
and more recently in \ci{Zarembo:2020tpf}  where it was found that (see \rf{126}) 
\be\la{2.10}
\vev{\mc W_1} \stackrel{\lambda\gg 1}{=}  w (\theta)\, W_0 \ , \ \ \ \ \ \ \ \ \ 
  \vev{\mc W_2}   \stackrel{\lambda\gg 1}{=} w (2\pi -\theta)\,  W_0 \ , \ \ \ \  \ \ \ \ 
W_0= N  {\sqrt{ 2\ov \pi} } \, \l^{-3/4} \, e^{\sqrt \l}   \ . \ee 
 $W_0$   is the leading   large $N$, strong coupling term in the SYM result in \rf{1.1}. 

\iffa 
\paragraph{Saddle point equations.}
The finite $N$ ``effective action'' is 
\be
S(a) = \sum_{a}\frac{8\pi^{2}N}{\lambda_{a}}\sum_{i}a_{ai}^{2}-\sum_{a}\sum_{i<j}\Big[\log(a_{ai}-a_{aj})^{2}+\log H^{2}(a_{ai}-a_{aj})\Big]+\sum_{ij}\log H^{2}(a_{1i}-a_{2j}).
\ee
Hence, taking a derivative with respect to  $a_{1k}$, 
\be
0 = \frac{16\pi^{2}N}{\lambda_{1}} a_{1k}-2\,\sum_{j\neq q}\Big[\frac{1}{a_{1k}-a_{1j}}-K(a_{1k}-a_{1j})\Big]-2\sum_{j\neq k} K(a_{1k}-a_{2j}),
\ee
which is equivalent to the first equation in (\ref{2.6}), using (\ref{2.5}). \footnote{Of course, the fact that we are considering the large $N$ limit is implicit in the 
assumption that the condition $S'(a)=0$, \ie  the saddle point condition without considering fluctuations, captures the solution of the matrix model.}
\fi

\section{Weak  coupling expansion  in the orbifold theory} %  and weak coupling}

%At large $N$, there is equivalence with $\mc N=4$ SYM  for 
 %$\lambda_{1}=\lambda_{2}$  as may  be  checked  also  in weak coupling expansions in  \cite{Mitev:2015oty}.
 
 Considering the orbifold theory case $\lambda_{1}=\lambda_{2}$ one can work  out the  weak-coupling expansion 
 of $\vev{\mc W}^{\rm orb}  $   by starting with the  integral representation \rf{2.1},\rf{2.3}  for  finite $N$. 
 It     may  be formally  written as a sum of  functions $ W_{\zeta_{3}} (\l, N),\   W_{\zeta_{5}} (\l, N), ..., $  multiplying  particular  products 
 of $\zeta_n=\zeta(n)$  values\foot{Similar expansions are found in other  similar $\N=2$ models,  cf. \cite{Mitev:2015oty,Galvagno:2020cgq}.}
 \be
\la{3.1}
%\vev{\mc W}  \equiv
 \vev{\mc W}^{\rm orb}  =  \vev{\mc W}^{\rm SYM}+  W(\l,N) \ , \qquad 
 W= \zeta_{3}\, W_{\zeta_{3}}\, +\zeta_{5}\, W_{\zeta_{5}}\,
+\zeta_{7}\, W_{\zeta_{7}}\ +\zeta_{3}^{2}\, W_{\zeta_{3}^{2}}\, +\cdots \ . 
\ee 
Here $\vev{\mc W}^{\rm SYM}$ is  given by \rf{1.8}  so that  $W_{\zeta_{3}}$, etc.,   scale as $1/N$  at large $N$. 
% $\zeta$-contributions start at order $1/N$ and, at weak coupling, have increasing powers of $\lambda$:
For small $\l$  one has 
\be
W_{\zeta_{3}} = \mc O(\lambda^{3}), \qquad
W_{\zeta_{5}} = \mc O(\lambda^{4}), \qquad 
W_{\zeta_{7}}=  \mc O(\lambda^{5}), \quad W_{\zeta_{3}^{2}} = \mc O(\lambda^{5}), \ \ \ \  \  {\rm etc.} 
\ee
%\subsection{Computing the $\zeta$-dependent corrections}
To compute these  functions starting with \rf{2.1}  let us  note that  using \rf{2.2},\rf{2.7}  we get 
%(using that $\sum_i a_{\aai} =0$)
%%%%%%%%%%%%%%%%%%%%%%%%%%
\ba
\la{3.7}
\log\, f  %=\sum_{a}\sum_{i<j}\log H^{2}(a_{ai}-a_{aj})-\sum_{ij}\log H^{2}(a_{1i}-a_{2j}) \\
&= \sum_{i,j}\Big[\frac{1}{2}\sum_{\aa}\log H^{2}(a_{\aai}-a_{\aa j})-\log H^{2}(a_{1i}-a_{2j})\Big] \\
%&= 2\sum_{n=1}^{\infty}\frac{(-1)^{n}}{n+1}\zeta_{2n+1}\sum_{ij}\Big[\frac{1}{2}\sum_{\aa}(a_{\aai}-a_{\aaj})^{2n+2}-(a_{1i}-a_{2j})^{2n+2}\Big] \\
%&= 2\sum_{n=1}^{\infty}\frac{(-1)^{n}}{n+1}\zeta_{2n+1}\sum_{k=0}^{2n+2} C^k_{n+2 } \sum_{ij}\Big[\frac{1}{2}\sum_{\aa}a_{\aai}^{k}a_{\aaj}^{2n+2-k}-a_{1i}^{k}a_{2j}^{2n+2-k}\Big] \\
%&= 2\sum_{n=1}^{\infty}\frac{(-1)^{n}}{n+1}\zeta_{2n+1}\sum_{k=0}^{2n+2}C^k_{n+2 } \Big[\frac{1}{2}\sum_{\aa}\tr\, a_{\aa}^{k}\,\tr\, a_{\aa}^{2n+2-k}-\tr\, a_{1}^{k}\, \tr\, a_{2}^{2n+2-k}\Big]\notag \\
&= 2\sum_{n=1}^{\infty}\Big(\frac{\lambda}{8\pi^{2}N}\Big)^{n+1}\frac{(-1)^{n}}{n+1}\zeta_{2n+1}\sum_{k=0}^{2n+2}
C^k_{n+2 } \Big[\frac{1}{2}\sum_{\aa}\tr A_{\aa}^{k}\,\tr A_{\aa}^{2n+2-k}-\tr A_{1}^{k}\, \tr A_{2}^{2n+2-k}\Big],  
\notag
\ea
where we defined $C^k_{n+2 } \equiv  (-1)^{k}  \binom{2n+2}{k} $, \ 
   $a_\aa ={\rm diag} ( a_{\aa 1}, ..., a_{\aa_N})$ (with $\tr\, a_\aa=0$)   and we 
    also introduced  the rescaled  matrices $A_\aa$   (appearing in the exponent in \rf{2.1})
\be\la{34}
A_{\aa} \equiv \sqrt{ \frac{8\pi^{2}N}{\lambda}}\, a_{\aa}  \ . 
\ee
%%%%%%%%%%%%%%%%%%%%%%%%%%%%%%%
\subsection{Direct perturbative expansion}
%%%%%%%%%%%%%%%%%%%%%%%%%%%%%%%
Separating different  $\zeta_n$ terms    we may write $f$ in \rf{3.7}  as
an  expansion in $\frac{\lambda}{8\pi^{2}N} =  \frac{ g^2_{\rm YM}}{8 \pi^2} $
\ba
f =& 1-3\,\zeta_{3}\,\Big(\frac{\lambda}{8\pi^{2}N}\Big)^{2}\,\big(T^{(1)}_{2,2}+T^{(2)}_{2,2}-2T^{(1)}_{2}T^{(2)}_{2}\big)\lp\ \ 
+\frac{10}{3}\zeta_{5}\,\Big(\frac{\lambda}{8\pi^{2}N}\Big)^{3}\,\big(3T^{(1)}_{2,4}-2T^{(1)}_{3,3}-3T^{(1)}_{4}T^{(2)}_{2}-3T^{(1)}_{2}T^{(2)}_{4}+3T^{(2)}_{2,4}-2T^{(2)}_{3,3}\big)+\cdots, \la{35}\\
&  T^{(\aa)}_{n_{1}, n_{2}, \dots, n_r} \equiv \tr A_{\aa}^{n_{1}}\, \tr A_{\aa}^{n_{2}}\cdots \tr A_{\aa}^{n_{r}} \ . \la{36}
\ea
 Using \rf{3.7}   and  computing \rf{2.3}  by first  integrating out 
the  $A_{2}$ dependence  with the help of
\be\la{37}
\vev{\tr A_{2}^{2}}_0 = \tfrac{N^{2}-1}{2}, \qquad 
\vev{(\tr A_{2}^{2})^{2}}_0  = \tfrac{N^{4}-1}{4},\qquad \cdots,
\ee
we obtain  for the coefficient functions in \rf{3.1} 
\ba
W_{\zeta_{3}} =& -3\,\Big(\frac{\lambda}{8\pi^{2}N}\Big)^{2}  \Big\langle {\tr\, e^{\sqrt\frac{\lambda}{2N}\, A_{1}}\,\big[
:(\tr A_{1}^{2})^{2}:+2\,:\tr A_{1}^{2}:\big]}\Big\rangle_0\ ,\la{38}
\\ 
W_{\zeta_{5}} =& \frac{5}{3N}\,\Big(\frac{\lambda}{8\pi^{2}N}\Big)^{3}  \Big\langle \tr\, e^{\sqrt\frac{\lambda}{2N}\, A_{1}}\,\Big[
9 (-2+3 N^2)\, :\tr A_{1}^{2}:+\frac{6 N^3 (6+N^2) }{18-6 N^2+N^4}\,:\tr A_{1}^{4}: \lp
+6 (-3+2 N^2) \,: (\tr A_{1}^{2})^{2}:
+6 N\, :\tr A_{1}^{2}\, \tr A_{1}^{4}:-4 N\,:(\tr A_{1}^{3})^{2}:
\Big]\Big\rangle_0\ , \ ...\ , \la{39}
\ea
where  
\be\la{000}
:\tr A_{1}^{2}: = \tr A_{1}^{2}-\tfrac{N^{2}-1}{2} \ , \ \ \  \ \ \ \ \   \vev{ :\tr A_{1}^{2}:}_0 =0\ , \qquad     {\rm etc.} \ee
% $ \vev{:A^2:}_{0}= \vev{A^2}_{0} - \vev{A}_0^2 $, etc.
This reduces the problem   to  evaluating correlators in one-matrix  ($A=A_1$) Gaussian model.

Computing  the Wilson loop correlator with normal ordered operators using  that for the $SU(N)$ SYM  case \ci{Drukker:2000rr}
\be
\la{3.17}
\WSYM = \langle{\tr \, e^{\sqrt\frac{\lambda}{2N}\, A_{1}}} \rangle_0= e^{\frac{\lambda}{8N}\big(1-\frac{1}{N}\big)}\, L_{N-1}^{1}\Big(-\frac{\lambda}{4N}\Big) \ , 
\ee
and  applying  the method described in Appendix~\ref{app:rec} (see also \cite{Beccaria:2020ykg}), we find 
 ($g\equiv g_{\rm YM}=\sqrt {\l \ov  N} $) %,  \ \     $\partial_\l \equiv {\partial \ov \partial_{\lambda}}$)
 {\small 
\ba
%W_{2} =
& \vev{\tr\, e^{\sqrt\frac{\lambda}{2N}\, A_{1}}:\tr A_{1}^{2}:}_0 = \frac{g}{2}\partial_{g} \WSYM = \lambda\partial_{\lambda} \WSYM, \la{88} \\
%W_{2,2} =
& \vev{\tr\, e^{\sqrt\frac{\lambda}{2N}\, A_{1}}\,:(\tr A_{1}^{2})^{2}:} = 
\Big(\frac{g^{2}}{4}\partial^{2}_{g}-\frac{g}{4}\partial_{g}\Big)\, \WSYM = \lambda^{2}\partial^{2}_{\lambda} \WSYM,\notag \\
%%%%%%%%%%%%
%W_{4} =
& \vev{\tr\, e^{\sqrt\frac{\lambda}{2N}\, A_{1}}:\tr A_{1}^{4}:}_0 =\Big[ \tfrac{\lambda  (-1+N) (1+N) }{16 N^3}+\tfrac{\lambda  
(\lambda +8 N^2-8 N^4) }{8 N^3}\,\partial_\l
%&\qquad \qquad \qquad \qquad \qquad \qquad  
-\tfrac{\lambda  (\lambda -12 N^2)}{N}\,\partial^{2}_\l +8 \lambda ^2 N \,\partial^{3}_\l \Big]\,\WSYM, \\
%%%%%%%%%%%%%%%%%%%%%%%%%%%%
%W_{2,4} =
& \vev{\tr\, e^{\sqrt\frac{\lambda}{2N}\, A_{1}}\,:\tr A_{1}^{2}\, \tr A_{1}^{4}:}_0 = \Big[-\tfrac{\lambda  (-1+N) (1+N) }{16 N^3}+\tfrac{\lambda  
(-1+N) (1+N) (\lambda +16 N^2) }{16 N^3}\,\partial _\l 
+\tfrac{\lambda  (\lambda ^2+8 \lambda  N^2-96 N^4-8 \lambda  N^4) }{8 N^3}\,\partial^{2}_\l \nonumber \\ 
& \qquad \qquad \qquad \qquad \qquad \qquad \ \ \ \   -\tfrac{\lambda ^2 (\lambda -12 N^2) }{N}\,\partial^{3}_\l +8 \lambda ^3 N\,\partial^{4}_\l  \Big]\,\WSYM, \\
%%%%%%%%%%%%%%%%%%%%%%%%%%%%%%
%W_{3,3} =
& \vev{\tr\, e^{\sqrt\frac{\lambda}{2N}\, A_{1}}\,:(\tr A_{1}^{3})^{2}:}_0 =\Big[ \tfrac{\lambda  (-1+N) (1+N) (-108+72 N^2-9 N^4+N^6) }{32 
N^3 (18-6 N^2+N^4)} \lp\qquad \qquad \qquad \qquad \qquad \qquad  
+\tfrac{\lambda  (-45 \lambda -432 N^2+42 \lambda  
N^2+612 N^4-7 \lambda  N^4-216 N^6+\lambda  N^6+10 N^8) }{8 N^3 (18-6 N^2+N^4)}\,\partial_\l \lp\qquad \qquad \qquad \qquad \qquad \qquad  
+\tfrac{\lambda  (18 \lambda ^2-432 \lambda  
N^2-6 \lambda ^2 N^2-5184 N^4-72 \lambda  N^4+\lambda ^2 N^4+2592 
N^6+12 \lambda  N^6-144 N^8-8 \lambda  N^8) }{8 N^3 
(18-6 N^2+N^4)}\,\partial^{2}_\l \lp\qquad \qquad \qquad \qquad \qquad \qquad  
-\tfrac{2 \lambda ^2 (18 \lambda +36 N^2-6 \lambda  
N^2-48 N^4+\lambda  N^4-4 N^6) }{N (18-6 N^2+N^4)}\,\partial^{3}_\l +8 
\lambda ^3 N \,\partial^{4}_\l \Big]\,\WSYM\ . \la{888}
\ea
}
Since   according to \rf{3.17}
\be
\la{1.11}
\vev{\mc W}^{\rm SYM} =\frac{2N}{\sqrt\lambda}I_{1}+\frac{1}{N}\frac{\lambda}{48}\Big(I_{2}-\frac{12}{\sqrt\lambda} I_{1}\Big)+\mc O\Big({1\ov N^{3}}\Big) \ , 
\qquad \qquad I_n \equiv I_n(\sqrt \lambda) \ , 
\ee
we find for \rf{38},\rf{39}
\ba
W_{\zeta_{3}} =& -3\Big(\frac{\lambda}{8\pi^{2}N}\Big)^{2}  \Big[N\frac{\sqrt\lambda}{2}\,I_{1}+\frac{1}{N}
\Big(\frac{\lambda(\lambda-48)}{192}\,I_{0}-\frac{\lambda^{3/2}}{24}\,I_{1}\Big)+\mc O\Big({1\ov N^{3}}\Big)\Big]
, \la{319} \\
%%%%%%%%%%%%
W_{\zeta_{5}} =& \Big(\frac{\lambda}{8\pi^{2}N}\Big)^{3}\Big[
\frac{45}{4} N^2 \sqrt{\lambda }\,  I_1\, 
+\frac{5}{128} \lambda  (-160+3 \lambda ) I_0  
-\frac{5}{64} \sqrt{\lambda } (80+9 \lambda ) 
I_1 +\mc O\Big({1\ov N^{2}}\Big)  \la{3.19}
\Big] \ . 
\ea
The expressions  \rf{319} and \rf{3.19} 
 generate  all terms proportional to $\zeta_{3}$ and $\zeta_{5}$ 
at leading and subleading order in $1/N$  in \rf{3.1} within the  {\it weak coupling} expansion. 
As expected, these  functions scale as $1/N$  for   large $N$, i.e. 
{\small \ba
\la{3.20}
W_{\zeta_{3}} = -\frac{1}{N}\,\frac{3}{2}\Big(\frac{\lambda}{8\pi^{2}}\Big)^{2}\,\sqrt\lambda\,I_{1}(\sqrt\lambda)+\mc O\Big(\frac{1}{N^{3}}\Big), \qquad 
%%%%%%%%%%%%
W_{\zeta_{5}} = \frac{1}{N}\,\frac{45}{4}\,\Big(\frac{\lambda}{8\pi^{2}}\Big)^{3}\,\sqrt\lambda\,I_{1}(\sqrt\lambda)+\mc O\Big(\frac{1}{N^{3}}\Big).
\ea  }
%where $W$ is certainly organized in powers of zeta numbers.
%%%%%%%%%%%%%%%%%
Similarly,   the  leading large $N$ terms  in  other  coefficient functions in \rf{3.1}  are given by 
%To gain further information, we coded the manipulations that we have illustrated in the previous cases, \ie for the $\zeta_{3}$ and $\zeta_{5}$ contributions. At leading order we find the following extension of (\ref{3.19})
{\small
\ba
\la{3.22}
W_{\zeta_{7}} =& \frac{1}{N}\,\Big(\frac{\lambda}{8\pi^{2}}\Big)^{4}\,\Big[-\frac{315}{4}\,\sqrt\lambda\,I_{1} 
+\frac{1}{N^{2}}\Big(
-\frac{105}{128} (-56+\lambda ) \lambda  I_0+\frac{105}{32} \sqrt{\lambda } (28+\lambda ) I_1
\Big)+\cdots\Big], \notag \\
%%%%%%%%%%%%%%%%%%%%%%%%%%%%%%%%%
W_{\zeta_{3}^{2}} =&  \frac{1}{N}\,\Big(\frac{\lambda}{8\pi^{2}}\Big)^{4}\,\Big[9\,\sqrt\lambda\,I_{1}
+\frac{1}{N^{2}}\Big(
\frac{3}{32} \lambda  (24+\lambda ) I_0-\frac{3}{16} 
(-24+\lambda ) \sqrt{\lambda } I_1
\Big)+\cdots\Big], \notag \\
%%%%%%%%%%%%%%%%%%%%%%%%%%%%%%%%%
W_{\zeta_{3}\zeta_{5}} =&  \frac{1}{N}\,\Big(\frac{\lambda}{8\pi^{2}}\Big)^{5}\,\Big[-150\,\sqrt\lambda\,I_{1}
+\frac{1}{N^{2}}\Big(
-\frac{5}{16} \lambda  (96+5 \lambda ) I_0+\frac{5}{16} (-288+\lambda ) \sqrt{\lambda } I_1
\Big)+\cdots\Big], \notag \\
%%%%%%%%%%%%%%%%%%%%%%%%%%%%%%%%%
W_{\zeta_{9}} =&  \frac{1}{N}\,\Big(\frac{\lambda}{8\pi^{2}}\Big)^{5}\,\Big[\frac{2205}{4}\,\sqrt\lambda\,I_{1}
+\frac{1}{N^{2}}\Big(
\frac{105}{128} \lambda  (-384+7 \lambda ) I_0 -\frac{105}{64} \sqrt{\lambda } (576+7 \lambda ) I_1 \Big)+\cdots\Big], \notag \\
%%%%%%%%%%%%%%%%%%%%%%%%%%%%%%%%%
W_{\zeta_{3}^{3}} =&  \frac{1}{N}\,\Big(\frac{\lambda}{8\pi^{2}}\Big)^{6}\,\Big[-54\,\sqrt\lambda\,I_{1}
+\frac{1}{N^{2}}\Big(
-\frac{9}{16} \lambda  (96+\lambda ) I_0-\frac{9}{4} 
\sqrt{\lambda } (96+\lambda ) I_1
\Big)+\cdots\Big],  \\
%%%%%%%%%%%%%%%%%%%%%%%%%%%%%%%%%
W_{\zeta_{5}^{2}} =&  \frac{1}{N}\,\Big(\frac{\lambda}{8\pi^{2}}\Big)^{6}\,\Big[675\,\sqrt\lambda\,I_{1}
+\frac{1}{N^{2}}\Big(
\frac{25}{32} \lambda  (80+9 \lambda ) I_0+\frac{25}{64} \sqrt{\lambda } (640+27 \lambda ) I_1\Big)+\cdots\Big], \notag \\
%%%%%%%%%%%%%%%%%%%%%%%%%%%%%%%%%
W_{\zeta_{3}\zeta_{7}} =&  \frac{1}{N}\,\Big(\frac{\lambda}{8\pi^{2}}\Big)^{6}\,\Big[\frac{2205}{2}\,\sqrt\lambda\,I_{1}
+\frac{1}{N^{2}}\Big(
\frac{105}{64} \lambda  (160+7 \lambda ) I_0+\frac{105}{16} \sqrt{\lambda } (160+3 \lambda ) I_1\Big)+\cdots\Big], \notag \\
%%%%%%%%%%%%%%%%%%%%%%%%%%%%%%%%%
W_{\zeta_{11}} =&  \frac{1}{N}\,\Big(\frac{\lambda}{8\pi^{2}}\Big)^{6}\,\Big[-\frac{31185}{8}\,\sqrt\lambda\,I_{1}
+\frac{1}{N^{2}}\Big(
-\frac{1155}{256} \lambda  (-448+9 \lambda ) I_0+8085 \sqrt{\lambda } I_1
\Big)+\cdots\Big]. \notag
\ea
}
%%%%%%%%%%%%%%%%%%%%%%%%%%%%%%%
\subsection{Leading non-planar  correction  and relation to free energy}
%%%%%%%%%%%%%%%%%%%%%%%%%%%%%%%
Remarkably, the dependence on $\lambda$  of  the leading term $\mc O(1/N)$ in the  $W$ functions 
 in  \rf{3.1}  
follows the same pattern, i.e. is proportional 
to the  Bessel function $I_{1}(\sqrt\lambda)$  that appears 
%MB1
in the leading order term 
in the SYM  expression  \rf{1.11}
\be
\la{4.1}
%\langle \mc W\rangle
W_{{\prod_{n=1}^{r}\zeta_{2n+1}^{k_{n}}}} = \frac{c_{k_{1}  \dots k_r}}{N}\,\Big(\frac{\lambda}{8\pi^{2}}\Big) ^{\sum_{n=1}^{r}k_{n}(n+1)}\,\sqrt\lambda\,I_{1}(\sqrt\lambda)+
\mc O\Big(\frac{1}{N^{3}}\Big)\ . 
\ee
%%%%%%%%%%%%%%%%%%%%%
%Since $\sqrt\lambda\,I_{1}(\sqrt\lambda) = \frac{\lambda}{2}+\cdots$, we can read the coefficients $c_{k_{1}, k_{2}, \dots}$ from a calculation that in practice does not involve
%the detailed structure of the Wilson loop. \footnote{Of course, it is only for the Wilson loop that (\ref{4.1}) has that precise structure.} 
The power of $\frac{\lambda}{8\pi^{2}}$ 
is  coming from the $\zeta_{2n+1}$ factors in \rf{3.7}   while  the extra factor of  $\lambda$ 
(from the Bessel  function  factor $\sqrt\lambda\,I_{1}(\sqrt\lambda) = \frac{\lambda}{2}+\cdots$)
 has its origin in the Wilson loop  operator insertion into the Gaussian matrix model integral  at the leading order in large $N$ (cf.  \rf{2.3}).
 In particular,  it comes   from the   $A^2_1$ term  in    ($\tr A_\aa=0$)
\be \la{423}
\tr\, e^ {\sqrt\frac{\lambda}{2N}\,A_{1}} = N+\frac{\lambda}{4N}\tr\,  A_{1}^{2}+\cdots. \ee
%the $\zeta$-corrections have the same Bessel function dependence on $\lambda$ as the leading order, up to a power of $\lambda$.
Separating this SYM  Bessel function factor,  the expression for the leading large $N$ term in $W=       \vev{\mc W}^{\rm orb}  -  \vev{\mc W}^{\rm SYM}$ in  \rf{3.1}   can be written in terms of the function $q(\lambda)$  defined  in (\ref{1.9}),\rf{x3}
\ba
\la{3.23}
q^{\rm orb} (\lambda) =& q^{\rm SYM}(\lambda)+\Delta q(\lambda), \qquad \qquad 
\Delta q(\lambda) = \l  \sum_{n=2}^{\infty} d_{n}\,\Big(\frac{\lambda}{8\pi^{2}}\Big)^{n} \ .   
\ea
Here $q^{\rm SYM}(\lambda)$ was given  in   (\ref{1.10})  and the coefficients $d_{n}$  are  found to be (cf.  \rf{3.22}) 
% (including also some higher order terms)  %read (pushing the calculation to higher order)
 \ba
d_{2} &\te = -\frac{3}{4}\,\zeta_{3}, \qquad 
d_{3} = \frac{45}{8}\,\zeta_{5}\,\qquad
d_{4} = -\frac{315}{8}\,\zeta_{7}+\frac{9}{2}\,\zeta_{3}^{2},\qquad 
d_{5} = -75\,\zeta_{3}\,\zeta_{5}+\frac{2205}{8}\,\zeta_{9}, \notag \\
d_{6} &\te = -27\,\zeta_{3}^{3}+\frac{675}{2}\,\zeta_{5}^{2}+\frac{2205}{4}\,\zeta_{3}\,\zeta_{7}- \frac{31185}{16}\,\zeta_{11}, \qquad 
d_{7} = 630\zeta_{3}^{2}\zeta_{5}-\frac{41895}{8}\,\zeta_{5}\zeta_{7}-3969\,\zeta_{3}\zeta_{9}+\frac{891891}{64}\,\zeta_{13}, \notag \\
d_{8} &\te = 
162\,\zeta_{3}^4
-4950\,\zeta_{3} \zeta_{5}^2
-4410\,\zeta_{3}^2 \zeta_{7}
+\frac{337365}{16}\,\zeta_{7}^2
+\frac{78435}{2}\,\zeta_{5} \zeta_{9}
+\frac{114345}{4}\,\zeta_{3} \zeta_{11}
-\frac{6441435}{64}\zeta_{15}, \la{322} \\
d_{9} &\te = 
-4860\,\zeta_{3}^3 \zeta_{5}
+\frac{25875}{2}\,\zeta_{5}^3
+69930\,\zeta_{3} \zeta_{5} \zeta_{7}
+30618\,\zeta_{3}^2 \zeta_{9}  -\frac{5190885}{16}\,\zeta_{7} \zeta_{9}\lp\ \ \  \ \ \te 
-\frac{4655475}{16}\,\zeta_{5} \zeta_{11}
-\frac{1656369}{8}\,\zeta_{3} \zeta_{13}
+\frac{46930455}{64}\,\zeta_{17}. \notag
\ea  
In general, starting from  the definition \rf{2.3}  of the Wilson loop expectation value \rf{2.3}  (e.g. for $\aa=1$),  plugging in 
  the expansion of  $f$ in \rf{35}  and  taking  $N$  large   we get, using  that  the  integration 
   over   the ``decoupled''  variable   $A_2$   gives an extra  $SU(N)$ SYM   factor  (cf. \rf{1.9},\rf{x3}  and \rf{211},\rf{34},\rf{3.17})
\ba \la{101}
N \to \infty: \ \ \ \ \ \ \  \vev{\mc W}^{\rm orb} &= \vev{\mc W}^{\rm SYM }  \Big[ 1 +  
\frac{\frac{\sqrt\lambda }{2N}I_{1}\,\vev{:\tr A_{1}^{2}:\, f}_0}{\vev{f}_0}
\frac{1}{\frac{2N}{\sqrt\lambda}I_{1}}+\mc O\Big(\frac{1}{N^{4}}\Big)\Big]
 \ , \\
\la{4.4}
\Delta q(\lambda) &=  \lim_{N\to\infty}  \frac{\lambda}{4}\,\frac{\vev{:\tr A_{1}^{2}:\,f}_0}{\vev{f}_0} =\lim_{N\to\infty}\frac{\lambda^{2}}{8}\frac{d}{d\lambda}\log \vev{f}_0\ .
\ea
%%%%%%%%%%%%%%%%%%%%%%%%%%%%%%%%%%%%%%%%%%%%%%%%%%%%
%\section{Relating $q(\lambda)$ and the orbifold free energy}
We used  \rf{2.2},\rf{3.7} to  represent $\Delta q(\lambda)$ in \rf{1.9},\rf{x3},\rf{3.23}  in terms of    the  Gaussian matrix model expectation value, 
%Splitting out the $\mc N=2$ corrections as in (\ref{3.23}), we have 
%where $f$ has been defined in (\ref{3.4}), see also (\ref{3.7}).
%Indeed, the partition function in  \rf{2.1}  that normalizes  the correlator in {2.3}  is $Z= \langle f \rangle$
   traded the insertion of $:\tr A_{1}^{2}+\tr A_{2}^{2}:$  for 
 the  application of $\lambda\partial_{\lambda}$ and used the $A_1\leftrightarrow A_2$ symmetry
of the integration measure. The expression \rf{4.4} 
is equivalent  to \rf{x2},\rf{1.16}  representing   $ \Delta q(\lambda)  $
in terms  of the free energy $F= -\log Z$ of the  orbifold theory (see \rf{211}). 

As a check, using \rf{3.7} one  can compute the leading terms in the expansion of $ \vev{f}_0$
\ba
\la{327}
\vev{f}_0 = { Z^{\rm orb} \ov Z_0} =&\te  1-\frac{3 (N^{2}-1)}{N^2}\zeta_{3}\Big(\frac{\lambda}{8\pi^{2}}\Big)^{2}+\frac{5 (N^{2}-1)(3 N^2-2)}{N^4} \zeta_{5} \Big(\frac{\lambda}{8\pi^{2}}\Big)^3 \lp\te \ \ \ \ 
+\Big[\frac{27 (N^{2}-1) (N^2+1) }{2 N^4}\zeta_{3}^2-\frac{105 (N^{2}-1) (3 N^4-4 N^2+2) }{4 
N^6}\zeta_{7}\Big] \Big(\frac{\lambda}{8\pi^{2}}\Big)^4+\cdots\notag 
\\    \stackrel{N \to \infty }{=}& \te  1-3\,\zeta_{3}\Big(\frac{\lambda}{8\pi^{2}}\Big)^{2}+15\zeta_{5}\, \Big(\frac{\lambda}{8\pi^{2}}\Big)^3 +\Big[\frac{27}{2} \zeta_{3}^2-\frac{315}{4}\zeta_{7}
\Big] \Big(\frac{\lambda}{8\pi^{2}}\Big)^4+\cdots.
\ea
Hence, for \rf{4.4} we get 
\ba
&\Delta q= \frac{\lambda^{2}}{8}\frac{d}{d\lambda}\lim_{N\to\infty}\log \vev{f}_0 \te = -\frac{3}{4}\zeta_{3}\,\lambda\, \Big(\frac{\lambda}{8\pi^{2}}\Big)^{2}+\frac{45}{8}\zeta_{5}\,\lambda\, \Big(\frac{\lambda}{8\pi^{2}}\Big)^{3}
+\Big[\frac{9}{2}\zeta_{3}^{2}-\frac{315}{8}\zeta_{7}\Big]\,\lambda\, \Big(\frac{\lambda}{8\pi^{2}}\Big)^{4}+\cdots,
\ea
which is in agreement with (\ref{3.23}),\rf{322}.

Thus the problem of computing $\Delta q$ in \rf{3.23}  is reduced to the calculation
of the large $N$ limit of the   free energy  of the $\mc N=2$  
 orbifold  theory (which is finite for $N\to\infty$ after the subtraction 
 of the planar  $\mc N=4$  SYM  term).
 This method is rather efficient as the   direct   computation of $\Delta q$  to higher orders without 
exploiting the  resummation of the $\l$ dependence  in the factor $\sqrt\lambda I_{1}(\sqrt\lambda)$
 would be prohibitively difficult. 
Using it  we  were able to  
%find  the next five  coefficients  in 
%MB1
push the calculation of the $\zeta_n$ expansion of $q(\lambda)$  up to $\mc O(\lambda^{20})$.
In particular, the next five terms beyond (\ref{3.23}) read
{\small
\ba
d_{10} =&\te\te  -972\,\zeta_{3}^5+55125\,\zeta_{3}^2 \zeta_{5}^2+33075\,\zeta_{3}^3 
\zeta_{7}-\frac{2197125}{8}\,\zeta_{5}^2 \zeta_{7}-\frac{3980025}{16}
\,\zeta_{3} \zeta_{7}^2  
-\frac{978075}{2}\,\zeta_{3} \zeta_{5} \zeta_{9} \lp\te \te +\frac{40730445}{32}\,\zeta_{9}^2-\frac{1715175}{8}\,\zeta_{3}^2 
\zeta_{11}+\frac{157224375}{64}\,\zeta_{7} \zeta_{11} 
+\frac{69015375}{32}\,\zeta_{5} \zeta_{13}
+\frac{96621525}{64}\,\zeta_{3} \zeta_{15}-\frac{1378048815}{256}\,\zeta_{19}, \notag \\
%%%%%%%%%%%%%%%%%%%%%%%%%%%%%%%%%%%%%%%%%%%%%%%%%
d_{11} =&\te  35640\,\zeta_{3}^4 \zeta_{5}-280500\,\zeta_{3} \zeta^3_5
-755370\,\zeta_{3}^2 \zeta_{5} \zeta_{7}+\frac{62375775}{32}\,\zeta_{5} \zeta_{7}^2
-224532\,\zeta_{3}^3 
\zeta_{9}+\frac{15401925}{8}\,\zeta_{5}^2 \zeta_{9}  \lp\te \te+\frac{6995835}{2}
\,\zeta_{3} \zeta_{7} \zeta_{9}+\frac{6887925}{2}\,\zeta_{3} \zeta_{5} \zeta_{11} 
-\frac{1252497015}{64}\,\zeta_{9} \zeta_{11}+\frac{6073353}{4}\,\zeta_{3}^2 \zeta_{13} \lp\te  \te  -\frac{2368300935}{128}
\,\zeta_{7} \zeta_{13}-\frac{2050523475}{128}\,\zeta_{5} \zeta_{15} 
-\frac{88646415}{8}\,\zeta_{3} \zeta_{17}+\frac{20364499155}{512}
\,\zeta_{21}, \notag \\
%%%%%%%%%%%%%%%%%%%%%%%%%%%%%%%%%%%%%%%%%%%%%%%%%
d_{12} =&\te 5832\,\zeta_{3}^6-526500\,\zeta_{3}^3 \zeta_{5}^2+\frac{1085625}{2}\,\zeta_{5}^4-238140\,\zeta_{3}^4 \zeta_{7}
+\frac{11618775}{2}\,\zeta_{3} \zeta_{5}^2 \zeta_{7} 
+\frac{10405395}{4}\,\zeta_{3}^2 \zeta_{7}^2-\frac{147697515}{32}\,\zeta_{7}^3 \lp\te \te  +5154030\,\zeta_{3}^2 
\zeta_{5} \zeta_{9}-\frac{109609605}{4}\,\zeta_{5} \zeta_{7} \zeta_{9} 
-\frac{98765163}{8}\,\zeta_{3} \zeta_{9}^2+\frac{3087315}{2}\,\zeta_{3}^3 \zeta_{11}-\frac{108709425}{8}\,
\zeta_{5}^2 \zeta_{11}\lp\te \te  -\frac{395748045}{16}\,\zeta_{3} \zeta_{7} 
\zeta_{11} 
+\frac{9743451795}{128}\,\zeta_{11}^2-\frac{196003665}{8}\,
\zeta_{3} \zeta_{5} \zeta_{13}+\frac{4773806037}{32}\,\zeta_{9} \zeta_{13} \lp\te \te 
-\frac{173918745}{16}\,\zeta_{3}^2 \zeta_{15}+\frac{17818765965}{128}\,\zeta_{7} \zeta_{15}
+\frac{1909927305}{16}\,\zeta_{5} \zeta_{17} 
+\frac{5236585497}{64}\,\zeta_{3} \zeta_{19}-\frac{151323893721}{512}\,\zeta_{23}, \notag \\
%%%%%%%%%%%%%%%%%%%%%%%%%%%%%%%%%%%%%%%%%%%%%%%%%
d_{13} =&\te-252720\,\zeta_{3}^5 \zeta_{5}+3919500\,\zeta_{3}^2 \zeta_{5}^3
+7076160\,\zeta_{3}^3 \zeta_{5} \zeta_{7}-15151500\,\zeta_{5}^3 
\zeta_{7}  
-\frac{80691975}{2}\,\zeta_{3} \zeta_{5} \zeta_{7}^2+1592136
\,\zeta_{3}^4 \zeta_{9}\lp\te \te
  -39864825\,\zeta_{3} \zeta_{5}^2 \zeta_{9} -35638785\,\zeta_{3}^2 \zeta_{7} \zeta_{9} 
+\frac{3118780665}{32}\,
\zeta_{7}^2 \zeta_{9}+96523245\,\zeta_{5} \zeta_{9}^2\lp\te  -\frac{71138925}{2}\,\zeta_{3}^2 \zeta_{5} \zeta_{11}
+\frac{1550674125}{8}\,\zeta_{5} \zeta_{7} \zeta_{11}+\frac{1400944545}{8}\,\zeta_{3} \zeta_{9} \zeta_{11}
-\frac{21532797}{2}\,\zeta_{3}^3 \zeta_{13}
+\frac{1549708875}{16}
\,\zeta_{5}^2 \zeta_{13}\lp\te  +\frac{2824276455}{16}\,\zeta_{3} \zeta_{7} 
\zeta_{13}-\frac{149972417595}{128}\,\zeta_{11} \zeta_{13}
+\frac{1408331925}{8}\,\zeta_{3} \zeta_{5} \zeta_{15}-\frac{145088958105}{128}\,\zeta_{9} \zeta_{15}
+\frac{314291835}{4}\,\zeta_{3}^2 \zeta_{17}\lp\te 
-\frac{67093614315}{64}\,\zeta_{7} \zeta_{17}
-\frac{7140798405}{8}\,\zeta_{5} \zeta_{19}
-\frac{38877680205}{64}\,\zeta_{3} \zeta_{21}
+\frac{1130016089475}{512}\,\zeta_{25}, \notag \\
%%%%%%%%%%%%%%%%%%%%%%%%%%%%%%%%%%%%%%%%%%%%%%%%%
d_{14} =&\te-34992\,\zeta_{3}^7+4592700\,\zeta_{3}^4 \zeta_{5}^2-14713125\,\zeta_{3} \zeta_{5}^4+1666980\,\zeta_{3}^5 \zeta_{7}
-\frac{159013575}{2}\,\zeta_{3}^2 \zeta_{5}^2 \zeta_{7}-\frac{95527215}{4}\,\zeta_{3}^3 \zeta_{7}^2\lp\te  +\frac{319787475}{2}\,
\zeta_{5}^2 \zeta_{7}^2
+\frac{752101245}{8}\,\zeta_{3} \zeta_{7}^3-47508930\,\zeta_{3}^3 \zeta_{5} \zeta_{9}+104792625\,\zeta_{5}^3 \zeta_{9}\lp\te 
+556189200\,\zeta_{3} \zeta_{5} \zeta_{7} \zeta_{9}+\frac{979410285}{8}\,\zeta_{3}^2 \zeta_{9}^2
-\frac{21984271155}{32}\,\zeta_{7} \zeta_{9}^2-\frac{21611205}{2}\,\zeta_{3}^4 \zeta_{11}+\frac{1105144425}{4}
\,\zeta_{3} \zeta_{5}^2 \zeta_{11}\lp\te 
+\frac{3946846365}{16}\,\zeta_{3}^2 
\zeta_{7} \zeta_{11}-\frac{5522257125}{8}\,\zeta_{7}^2 \zeta_{11}-\frac{10943977275}{8}\,\zeta_{5} \zeta_{9} \zeta_{11}
-\frac{39833719695}{64}\,\zeta_{3} \zeta_{11}^2+\frac{1990403415}{8}\,\zeta_{3}^2 \zeta_{5} 
\zeta_{13}\lp\te  -\frac{44287988745}{32}\,\zeta_{5} \zeta_{7} \zeta_{13}
-\frac{20042115093}{16}\,\zeta_{3} \zeta_{9} \zeta_{13}+\frac{581635463409}{128}\,\zeta_{13}^2
+\frac{1217431215}{16}\,\zeta_{3}^3 \zeta_{15}
-\frac{22305908625}{32}\,\zeta_{5}^2 \zeta_{15}\lp\te  -\frac{20343808485}{16}\,\zeta_{3} \zeta_{7} \zeta_{15}
+\frac{2297131261425}{256}\,\zeta_{11} \zeta_{15} 
-\frac{10210455255}{8}\,\zeta_{3} \zeta_{5} \zeta_{17}
+\frac{68849861535}{8}\,\zeta_{9} \zeta_{17}
-\frac{36656098479}{64}\,\zeta_{3}^2 \zeta_{19}\lp\te 
+\frac{2024629961445}{256}\,\zeta_{7} \zeta_{19}
+\frac{214325672925}{32}\,\zeta_{5} \zeta_{21}
+\frac{1160149851861}{256}\,\zeta_{3} \zeta_{23}
-\frac{16950241342125}{1024}\,\zeta_{27}. \la{329}
\ea
}

\def \z {\zeta}

\section{Resummation of   particular   transcendental  contributions to  $\vev{\mc W}^{\rm orb}$ } %  and their strong coupling limit }

In an attempt to  shed light on the structure of strong coupling limit of $\vev{\mc W}^{\rm orb}$ 
%(at first subleading order in $1/N$, i.e.  in $\Delta q (\l)$ in \rf{1.9})
one may try to  consider  separate terms in the transcendental part of \rf{3.1}, resum their weak coupling expansion and then expand  at strong coupling. As we shall see,  this procedure  will not give the correct strong-coupling limit of $\vev{\mc W}^{\rm orb}$: 
  the strong-coupling   asymptotics of different functions $W_{\zeta_n^k...}$  will be different. That means  that   all such terms  should  first  be summed  up before taking  the large $\l$   limit. 

\iffa 
 reach the strong coupling regime, we analyze in this section partial resummations
of classes of $\zeta$-monomials. However, it must be stressed that 
this approach is not expected to give correct strong coupling results unless one resums all terms. Nevertheless, to clarify the structure of the
function $q(\lambda)$, it may be interesting to gather information about special contributions.
\fi

As we shall show in Appendix \ref{B}  the terms  in \rf{3.1},\rf{4.1}    which are  proportional to the single  $\zeta_{2n+1}$ 
  have the following coefficient functions %(expanded at large $N$) 
\be \la{41}
W_{\z_{2n+1} }= \frac{2}{N}\Big(\frac{\lambda}{8\pi^{2}}\Big)^{n+1}\frac{(-1)^{n}}{n+1}
\frac{  2^{3n-1}\, 3 \ \Gamma(n+\frac{1}{2})\Gamma(n+\frac{3}{2})}{(n+2)\,\pi\,\Gamma(n)\Gamma(n+1)}
\sqrt\lambda\,I_{1}(\sqrt\lambda)+\mc O\Big(\frac{1}{N^{3}}\Big).
\ee
Summing   all such    terms  in  $\vev{\mc W}^{\rm orb}$ in \rf{3.1}, i.e. $ W\big|_{\rm  \zeta} = \sum^\infty_{n=1}  W_{\z_{2n+1} }$,    we then get 
the corresponding  contribution to $q^{\rm orb}$  or to  $\Delta q$  in \rf{x3},\rf{4.4} 
\be\la{42}
 \Delta q(\lambda)\Big|_{\rm  \zeta} =\sum_{n=1}^{\infty}(-1)^{n}\,\lambda\,\Big(\frac{\lambda}{8\pi^{2}}\Big)^{n+1}
%\frac{3}{\pi}
\frac{2^{3n-1}\, 3 \ \Gamma(n+\frac{1}{2})\Gamma(n+\frac{3}{2})}{\pi \Gamma(n)\Gamma(n+3)}\,\zeta_{2n+1}\ . 
\ee
%in agreement with (\ref{5.1}). 
We can resum  this series by noting  that 
\be\la{4.3}
\zeta_{2n+1} \equiv \zeta(2 n+1) =  \frac{1}{(2n)!}\int_{0}^{\infty}dt\ g(t) \ {t^{2n}} \ , \qquad \ \ \ 
 g(t) = \frac{1}{e^{t}-1} \ . 
\ee
This gives ($J_n$ are Bessel   functions)
 \ba\la{44}
&\Delta q(\lambda)\Big|_{\rm \zeta}   = \frac{\lambda^{2}}{16\pi}\,
\int_{0}^{\infty}dt\,  g(t) \,  f(t \sqrt \lambda)\ ,  %\ ,\qquad x = \sqrt\lambda \ , 
\\  &
 f(t) = 3 \,  [J_0(t )]^2-\frac{12\,J_0(t)\, J_1(t)}{t}-\frac{3 (t^2-4) \,
[J_1(t)]^2}{t^2}\ .
\ea
Using  the  properties   of the Mellin transform\foot{Defining  the Mellin transform 
$ \widetilde{f}(s) = \int_0^\infty dx \,  x^{s-1}\,f(x)\,$ 
 and considering  the convolution 
$(f\star g) (x)= \int_0^\infty dt \,  f(t\,x)\,g(t)$  we have   $
(\widetilde{f\star g})(s)  = \widetilde{f}(s)\,\widetilde{g}(1-s).
$
Let $\alpha< s< \beta$ be the fundamental strip of analyticity of $\widetilde{f}(s) $.
The asymptotic expansion of $f(x)$ for $x\to\infty$ is obtained by 
looking at the poles of $\widetilde{f}(s)$ in the region $s\ge \beta$. Then    the pole 
$
\frac{1}{(s-s_0)^N} $ in the Mellin transform  leads to  the term  $ \frac{(-1)^N}{(N-1)!}\,\frac{1}{x^{s_0}}\,\log^{N-1} x
$ in the original function. 
 }
   %described in  Appendix  \ref{sec:Mellin} 
we find   that the   large $\lambda$ asymptotics of \rf{44} is  % expansion  of this function is 
\be\la{46}
%\left. q(\lambda)\right|_{\rm single-\zeta} 
\Delta q(\lambda)\Big|_{\rm \zeta}  \stackrel{\l\gg 1}{=}  \frac{\lambda^{2}}{16\pi^2}\,\Big[-\frac{3}{8}+\lambda^{-1/2}
-\frac{9}{4}\,\zeta_{3}\,\lambda^{-3/2}+\cdots\Big].
\ee
Similarly, we may consider  all terms  in \rf{3.1}  proportional to $\zeta_{3}\zeta_{2n+1}$  with $n > 1$ (see Appendix \ref{B}). 
%\subsection{$\zeta_{3}\zeta_{2n+1}$ terms}
%%%%%%%%%%%%%%%%%%%%%%%%%%%%%%%%%%%%%%%
We get  the following analog of   \rf{42}
%The final result is 
\be
\la{47}
\Delta q(\lambda)\Big|_{\zeta_{3}\zeta} =-\sum_{n=2}^{\infty} (-1)^{n}\frac{3}{\pi} \frac{2^{3n+3}\,n\,[\Gamma(n+\frac{3}{2})]^{2}}{\Gamma(2+n)\Gamma(3+n)}\,\lambda\,\Big(\frac{\lambda}{8\pi^{2}}\Big)^{n+3}\,\zeta_{3}\,\zeta_{2n+1}\ . 
\ee
Summing this series   as in \rf{4.3},\rf{44}   we get 
%v2
\ba
\Delta q(\lambda)\Big|_{\zeta_{3}\zeta} 
&= -\tfrac{9\zeta_{3}^{2}}{4096\pi^{8}}\lambda^{5}+\tfrac{3\,\zeta_{3}\,}{64\pi^{5}}\lambda^{4} \, \int_{0}^{\infty}\frac{dt}{e^{2\pi t}-1}\
\hat f(t\sqrt\lambda),\la{48} \\
 \hat f(t) &= 
-[J_0(t)]^2+\frac{5 J_0(t) J_1(t)}{t}
+\frac{(t^2-6 ) [J_1(t)]^2}{t^2}.
\ea
Expanding at large $\lambda$ here  gives a  different asymptotics than in \rf{46}
%v2
\be\la{4.10}
\Delta q(\lambda)\Big|_{\zeta_{3}\zeta}   \stackrel{\l\gg 1}{=}  -\tfrac{9\zeta_{3}^{2}}{4096\pi^{8}}\lambda^{5}+\tfrac{3\,\zeta_{3}\,
}{1024\pi^{6}} \lambda^{4}
\Big[1-\tfrac{4\pi^2}{3}\lambda^{-1}+{24\zeta_{3}}\lambda^{-3/2}+\cdots\Big]
\ee
%The asymptotic exponent is $5$ and this is not compatible with the numerical analysis presented in the next section. 
%This shows that partial resummations of classes of $\zeta$ monomials is unable to capture the behaviour of the non-perturbative $q(\lambda)$ function.
As another example one may consider  all the  terms in $W$  involving only powers of $\z_3$. 
The resulting contribution takes a simple form  (see Appendix \ref{B})
\be
\la{411}
\Delta q(\lambda)\Big|_{\sum_n \zeta^n_{3}} = \sum_{n=1}^{\infty} \frac{1}{4}(-1)^{n}2^{n-1}3^{n}\,\lambda\Big(\frac{\lambda}{8\pi^{2}}\Big)^{2n}\,\zeta_{3}^{n} = -\frac{3 \lambda ^3 \zeta_{3}}{8 (32 \pi ^4+3 \lambda ^2 \zeta_{3})}   \stackrel{\l\gg 1}{=} 
 -\frac{\lambda}{8}+\cdots \ , 
\ee
with the large $\l$  asymptotics being again different from \rf{46} and \rf{4.10}.

Using  the general method described in Appendix \ref{B}   one is  able to  generalize \rf{411}
 to the sum of 
all  contributions   involving arbitrary powers of $\z_3$ and $\z_5$   and  also of $\z_7$
\ba
& \Delta q(\lambda)\Big|_{\sum_{n,m} \z^n_3\z_5^m} =  
\frac{3\l (-2 t_3+15 t_5-50 t_3 t_5+450 t_5^2+750 t_5^3)}{8 (1+10 t_5) 
(1+6 t_3-40 t_5-50 t_5^2)} \Big|_{t_{3} =  \zeta_{3}(\frac{\lambda}{8\pi^{2}})^{2}, \ 
t_{5} = \zeta_{5}(\frac{\lambda}{8\pi^{2}})^{3}  } %\notag \\
  \stackrel{\l\gg 1}{=}  -\frac{9}{16}\lambda + ... 
   \la{412}\\  
   %\ea
 % Similarly, terms will all possible powers of $\z_3, \z_5$ and $\z_7$ are summed by 
%\ba
&\Delta q(\lambda)\Big|_{\sum_{n,m,k} \z^n_3\z_5^m\z^k_7} 
= 
-\lambda\,\frac{N(t_{3}, t_{5}, t_{7})}{D(t_{3}, t_{5}, t_{7})} \Big|_{  t_{3} =  \zeta_{3}(\frac{\lambda}{8\pi^{2}})^{2}, \ 
t_{5} = \zeta_{5}(\frac{\lambda}{8\pi^{2}})^{3} ,\   t_{7} = \zeta_{7}(\frac{\lambda}{8\pi^{2}})^{4} } 
\stackrel{\l\gg 1}{=}  -\frac{5}{4}\lambda + ...  \ , \la{413} \\
&N(t_{3}, t_{5}, t_{7}) = 
-3 (32 t_3-240 t_5+800 t_3 t_5-7200 t_5^2-12000 t_5^3+1680 t_7-8400 
t_3 t_7+135240 t_5 t_7  \nonumber \\
&\qquad\qquad \qquad +25200 t_3 t_5 t_7+252000 t_5^2 t_7-635040 
t_7^2-323400 t_3 t_7^2-1670900 t_5 t_7^2+343000 t_5^2 t_7^2  \nonumber \\
&\qquad\qquad\qquad
+5227320 
t_7^3-720300 t_3 t_7^3-3344250 t_5 t_7^3+18727800 t_7^4+6302625 t_7^5),
\notag\\
&D(t_{3}, t_{5}, t_{7}) = 
4 (4+40 t_5-420 t_7-735 t_7^2) (-8-48 t_3+320 t_5+400 t_5^2-2100 
t_7-840 t_3 t_7    \nonumber \\
&\qquad\qquad\qquad
-3360 t_5 t_7+16170 t_7^2+5145 t_7^3) \ . 
\ea
Comparing \rf{411},\rf{412} and \rf{413}    suggests that the  strong  coupling limit  of the sum of monomials involving powers of the first  $k$    constants 
 $\z_3, \z_5, ..., \z_{2k+1}$  should be  (cf. \rf{411},\rf{412},\rf{413})
 \be
\Delta q(\lambda)\Big|_{\sum_{n_1,...,n_k } \z^{n_1}_{3}...\z^{n_k}_{2k+1}} 
\stackrel{\l\gg 1}{=}  -\frac{1}{16}(k+1) ( 2 k-1) \lambda + ...  \ , \qquad \ \ \  k=1,2, 3, ... \la{415} \\
\ee
Since  the coefficient in \rf{415} grows with $k$,  summing  up  such   contributions {\it after} taking the large $\l$ limit 
would not give a meaningful result.

%%%%%%%%%%%%%%%%%%%%%%%%%%%%%%%%%%%%%%%%

\section{Weak coupling expansion  for   non-symmetric quiver } %[Appebdix?]}
%%%%%%%%%%%%%%%%%%%%%%%

Let us now discuss  the  expectation value of the Wilson loops \rf{0},\rf{2.3}  in the 
case of the $SU(N) \times SU(N)$ quiver  for    unequal couplings 
$\l_1 \not= \l_2$.
We shall   consider for definiteness  $\vev{\mc W}\equiv \vev{\mc W_{1}}$.
Setting
\be
\lambda_{2}=\rho\,\lambda_{1} \ , \qquad \la{51}
\ee
the generalization of  (\ref{3.7})  will read 
\ba
\log \,  f =&2\sum_{n=1}^{\infty}\Big(\frac{\lambda_{1}}{8\pi^{2}N}\Big)^{n+1}\frac{(-1)^{n}}{n+1}\zeta_{2n+1}\sum_{k=0}^{2n+2}
(-1)^{k}  \binom{2n+2}{k}\lp
\times \Big[
\frac{1}{2}\sum_{a}\tr A_{1}^{k}\,\tr A_{1}^{2n+2-k}
+\rho^{n+1}\frac{1}{2}\tr A_{2}^{k}\,\tr A_{2}^{2n+2-k}
-\rho^{n+1-\frac{k}{2}}\,\tr A_{1}^{k}\tr A_{2}^{2n+2-k}\Big]. 
\ea
It is then  straightforward  to compute the  large $N$ expansion of the 
 coefficient functions of the  $\zeta_n$-monomial contributions to 
$\vev{\mc W}$ in the analog of \rf{3.1}. 
The first of them % is the contribution proportional to $\zeta_{3}$ and a short calculation
 that  generalizes  (\ref{319})  is 
\ba
\la{7.3}
W_{\zeta_{3}} &= \Big(\frac{\lambda_{1}}{8\pi^{2}N}\Big)^{2}  \Big[
3 N^3 \,(\rho-1)\,I_2(\sqrt{\lambda_{1} })\,- N \Big( \, 3  (\rho-1)(1 +  \frac{1}{8} \lambda_{1})\, 
I_0(\sqrt{\lambda_{1} })\lp\qquad \qquad \qquad  
- \frac{
(192 + \lambda_{1} ^2) (\rho-1) -48 \lambda_{1}   }{32 
\sqrt{\lambda_{1} }}I_1(\sqrt{\lambda_{1} })\Big)
+\mc O\big({1\ov N}\big)\Big].
\ea
The planar (order $N$) contribution here agrees with the  $N\to \infty$  part of the $\mc N=4$  SYM  result in \rf{1.1} 
 expressed in terms of  the effective
coupling   % introduced in  
%v2a
\cite{Pomoni:2013poa,Mitev:2014yba,Mitev:2015oty} ($g^{2}$ there  is $\frac{\lambda_1}{16\pi^{2}}$)\footnote{
 %MB1
 In the  case of the $\l_1\not= \l_2$  quiver 
 the   weak coupling expansion in the planar limit  was  analysed  also in  \cite{Pini:2017ouj} 
and \cite{Fiol:2020ojn}.} 
\ba
\frac{2N}{\sqrt\lambda_{\rm eff}}I_{1}(\sqrt\lambda_{\rm eff}) = &
\frac{2N}{\sqrt\lambda_{1}}I_{1}(\sqrt\lambda_{1})+3N\,\zeta_{3}\,(\rho-1)
\Big(\frac{\lambda_{1}}{8\pi^{2}}\Big)^{2}
\,I_{2}(\sqrt\lambda_{1})+\cdots \ , \\
\la{7.4}
\lambda_{\rm eff} = &\lambda_{1}+  12 \,\zeta_{3}\,(\rho-1)\,\frac{\lambda_{1}^{3}}{(16\pi^{2})^{2}}+\cdots \ .
\ea
%in agreement with the planar part in (\ref{7.3}). 
Note that the  subleading terms  in \rf{7.3}  proportional to $\rho -1$ are   similarly captured by the SYM term  
if we  modify \rf{7.4}  as  
\be\la{75}
\lambda_{\rm eff} = \lambda_{1}+12\,\zeta_{3}\,(\rho-1)\,\Big(1-\frac{1}{N^{2}}\Big)\,\frac{\lambda_{1}^{3}}{(16\pi^{2})^{2}}+...
% \text{higher monomials}.
\ee
One can  also  find the analog of $W_{\zeta_{5}}$  in \rf{3.1},\rf{3.19}  and the $\rho-1$ terms there  can be generated  from the SYM expression by the  replacement $\l\to \lambda_{\rm eff} $  generalizing \rf{75}\foot{The $N\to \infty$ limit of this expression is 
in agreement  with Eq. (35) in  \cite{Mitev:2015oty}.}
\ba
\lambda_{\rm eff} = & \te \lambda_{1}+12\,\zeta_{3}\,(\rho-1)\,\Big(1-\frac{1}{N^{2}}\Big)\,\frac{\lambda_{1}^{3}}{(16\pi^{2})^{2}} \lp 
\qquad\te  - \zeta_{5}\,(\rho-1)\,\Big[40(\rho+3)\Big(1-\frac{7}{2N^{2}}\Big)\frac{\lambda_{1}^{4}}{(16\pi^{2})^{3}}+\frac{40\pi^{2}}{3}\Big(1-\frac{2(13+3\rho)}{N^{2}}\Big)\frac{\lambda_{1}^{5}}{(16\pi^{2})^{4}} \lp\qquad \qquad \qquad\te  \ \ \ \ \ \
+\frac{64\pi^{4}}{9}\Big(1+\frac{74+15\rho}{4N^{2}}\Big)\,\frac{\lambda_{1}^{6}}{(16\pi^{2})^{5}}+\mc O(\lambda_{1}^{7})\Big]+...\ . %\text{higher monomials}.
\ea
This   suggests that  some  essential features  of the weak coupling expansion of the Wilson loop in the non-symmetric quiver case are already captured by the  orbifold case ($\rho=1$)  discussed  above.

%%%%%%%%%%%%%%%%%%%%%%%%%%%%%%%%%%%%%%%%%%%%%%%%%%%%%%%%%
\section{Numerical analysis of  the  quiver matrix model}
%%%%%%%%%%%%%%%%%%%%%%%%%

One may  try to  compute the Wilson loop  \rf{0} numerically  at finite $N$  and $\l$ 
 by starting with the matrix model  representation \rf{2.1},\rf{2.3}. 
 While this is a finite dimensional integral,  %in $\mc O(N)$ variables,
   the fact that are interested in  the limit $N\gg 1$ makes  the numerical integration hard.
 At the same time, we expect that, in the large $N$ limit, the relevant subset of the integration domain reduces to  
 a neighbourhood of the saddle point solution. %of the equations of motion.
  This problem  is completely analogous to the one in the 
  lattice field theory   (where   one computes  
   quantum corrections by numerical path integration with $N\sim \hbar^{-1}$) and may thus  be 
 dealt with by   the same Monte Carlo (MC) methods  (see,  for  instance,  \cite{Rothe:1992nt}).  

\subsection{Orbifold theory} %The $\mathbb{Z}_{2}$ symmetric point}

We analysed the  Wilson loop expectation value \rf{3.1} in the orbifold  case  by means of a 
% at finite $N$ and $\lambda$ by a 
 Metropolis-Hastings Monte Carlo  simulation \cite{Binder:2002} of the integral \rf{2.1},  a  robust approach
 that does not require fine tuning.\foot{For  other papers using  MC methods
in matrix models  see,  e.g.,    \cite{Ambjorn:2000dj,Ambjorn:2000dx,Hanada:2012si, Sasakura:2019hql,
Sasakura:2020jis,Tanwar:2020fuv}  and  section 6.5 of \cite{Joseph:2019zer}.
In the explored region of parameter space  the Metropolis-Hastings   algorithm turned out to be faster than the Hybrid Monte Carlo one, taking into account
autocorrelation.  The latter algorithm 
%(see e.g. \cite{Ambjorn:2000dj,Ambjorn:2000dx,Hanada:2012si, Sasakura:2019hql,
%Sasakura:2020jis,Tanwar:2020fuv}  and  section 6.5 of \cite{Joseph:2019zer})
is expected to be preferable at higher $N$ and possibly more 
efficient for large-scale simulations  which are beyond the scope of the present analysis.} 
 %%%%
 
 Given a configuration $X$ of the eigenvalues $a_{\text{a}i}$  corresponding to the two $SU(N)$  groups, 
the matrix integral (\ref{2.1}) weights each observable $\mc O(X)$, like the Wilson loop,  with a 
positive number $\exp(-S)$ where $S=S(X)$   corresponds to  the total integrand in (\ref{2.1}) including 
the Vandermonde factor. 
A Markov chain obeying detailed balance is built by making a local variation of $X\to X'$ and accepting the new configuration if $S(X')<S(X)$ or, in the case $S(X')>S(X)$,
with probability $e^{S(X)-S(X')}$.  We tuned the local changes of configuration in order to have an acceptance probability around 50-60\% which is a reasonable choice. 
Iterating this procedure produces a sequence $\{X_{n}\}$ of configurations distributed according to $\exp(-S)$ and one can measure the quantum expectation value as
the ensemble average
$\vev{\mc O} = \lim_{n\to \infty}\frac{1}{n}\sum_{m=1}^{n}\mc O(X_{m})$.  The sequence $\{X_{n}\}$ is correlated and its autocorrelation time has been measured at each data point 
and taken into account in the estimate of  an  error 
in this MC evaluation of $\vev{\mc O}$. \footnote{As is well known (see, for instance, \cite{sokal1997monte}), 
 denoting by $\vev{\cdots}_{\rm MC}$ the average over MC realisations and assuming an exponential autocorrelation 
for the measurements  $\mc O_{n}=\mc O(X_{n})$, \ie $\vev{\mc O_{n}\mc O_{m}}_{\rm MC} = \sigma_{\mc O}^{2}\,e^{-|n-m|/\tau_{\mc O}}$,
the variance of the expectation value estimator $\frac{1}{n}\sum_{m=1}^{n}\mc O(X_{m})$ is $\frac{1}{n^{2}}\sum^n_{m, k}\vev{\mc O_{m}\mc O_{k}}_{\rm MC} \sim \frac{2\tau_{\mc O}+1}{n}\sigma_{\mc O}^{2}$ showing that the effective number
of decorrelated measurements is roughly $n_{\rm decorr} = n/(2\tau_{\mc O}+1)$ which is the  factor entering the standard deviation of measurements $\sigma = \sigma_{\mc O}/\sqrt{n_{\rm decorr}}$. }

For each  value of $\lambda$, we ran our code at various values of $N$ and fitted the Wilson loop measurements
in order to extract the function $q^{\rm orb}(\lambda)$  \rf{3.23} 
that  governs the leading  non-planar correction in \rf{1.9}. The procedure is illustrated in  Fig. \ref{fig:Nfit} (left)
at the  value $\lambda=1$.  Fig. \ref{fig:Nfit} (right)  shows the histogram of measurements of the Wilson loop 
at $\lambda=200$, $N=20$, as a sample point.

%\paragraph{Weak coupling check}

To provide  non-trivial checks  %of the correctness 
of the numerical code  we considered the Wilson loop at $\lambda=1$
which is  a relatively  weak coupling.  %region.
 From (\ref{3.23}),\rf{322}  we see that  for this value the  leading $\zeta_n$ contributions  are negligible
 and we  may assume  that the same is true also  for higher order  contributions.  Then 
%thus we may assume that also additional higher $\zeta$-terms does not contribute at this value $\lambda=1$. 
%We then have 
\be\la{61}
q^{\rm orb}(1) = -0.122(2)\ ,
\ee
where the error is an estimate of the systematic error determined by including or not the
the  contribution of the $\zeta_n$ terms 
 explicitly 
 computed   above. 
 The extrapolated slope from the  finite $N$  MC simulations 
at $\lambda=1$ shown in Fig.~\ref{fig:Nfit} (left)   gives 
\be\la{62}
\text{MC}:\qquad q^{\rm orb}(1) = -0.12(1)\ ,
\ee
which is thus consistent with the analytic estimate \rf{61}.

%\paragraph{Moderate coupling}

To compare  results at  higher  values  of  $\l$ we need
to resum  %some kind of resummation of 
the perturbative expansion of $\Delta q(\lambda)$ in \rf{3.23},\rf{322},\rf{329}. 
We performed a Borel-Pad\'e resummation for  values of $\lambda$ up to 50, 
see Fig.~\ref{fig:pade}. The red line there  is the perturbative series which is expected to converge for 
$|\lambda| < \pi^{2}$ with  partial sums blowing up beyond that value.
\footnote{This is the radius of convergence of perturbative expansion in SYM theory
in the planar limit. 
Its origin  may be attributed to  the form of the single-magnon dispersion relation, which follows from superconformal symmetry 
\cite{Beisert:2004hm,Beisert:2006ez} 
and  it may also  be  found  using   the quantum algebraic curve approach 
 \cite{Gromov:2017blm}.  That 
 such a singularity is also present in  the $\mc N = 2$ theories was  first noticed in the mass-deformed $\mc N=2^{*}$ theory 
case  in \cite{Russo:2013qaa}. 
}
  The green line is the $[7/6]$ Pad\'e approximant of the Borel improved series, while 
the blue line is the Borel transform,  which is thus  in good agreement with the numerical  data points.

%\paragraph{Large coupling}

At   higher  values  of $\lambda$ we found similar extrapolations in $1/N^{2}$. 
 In Fig.~\ref{fig:res} (left) we show the intercept of the extrapolation which is expected to be 1, see (\ref{3.22}). 
This is a measure of the 
systematic error associated with the fit of the $N$ dependence. It increases with $\lambda$ and we increased the maximal  $N$ in order to keep it 
below the 0.2\% level.\foot{Let us note that 
%AT
 our numerical analysis is  in a region of values  of  $(\l, N)$ expected to be free from the instanton corrections 
which are weighted by the typical $\exp(-8\pi^{2}N/\lambda)$ factors,  at least up to instanton moduli space volume corrections \cite{Russo:2013kea}.}

The resulting function $q^{\rm orb}(\lambda)$ computed for up  to $\lambda=450$ is shown
in Fig.~ \ref{fig:res} (right).  In the $SU(N)$   $\mc N=4$  SYM  theory, we know from \rf{1.10}  that  at strong coupling 
$q^{\rm SYM}(\lambda)  = \frac{1}{96}\lambda^{3/2}+...$ 
which is   valid  with high accuracy already at $\lambda\gtrsim 20$.
In the orbifold theory we find that   $q^{\rm orb}(\lambda)$ is negative with a clear 
 bending at large $\l$ suggesting an asymptotic behaviour  
\be
\la{8.3}
q^{\rm orb}(\lambda)\sim -  \lambda^{\eta} \ ,\qquad \qquad  \eta>1 \ . 
\ee 
 The best fit of the blue data points in Fig.~\ref{fig:res} (right)
gives $\eta=1.49(2)$ where the conservative error estimate includes statistics as well as the systematic effects due to the
choice of fitting window. We estimated the latter by dropping some of the data points at smaller  values of $\lambda$. 
This exponent is  still to be taken with some caution since it is hard to say whether we are already in the asymptotic $\lambda\to \infty$ region but it appears to match the string theory prediction  in \rf{x1}  (see  also \rf{x7},\rf{xx7}).

\begin{figure}
\begin{center}
\includegraphics[width=0.6\textwidth]{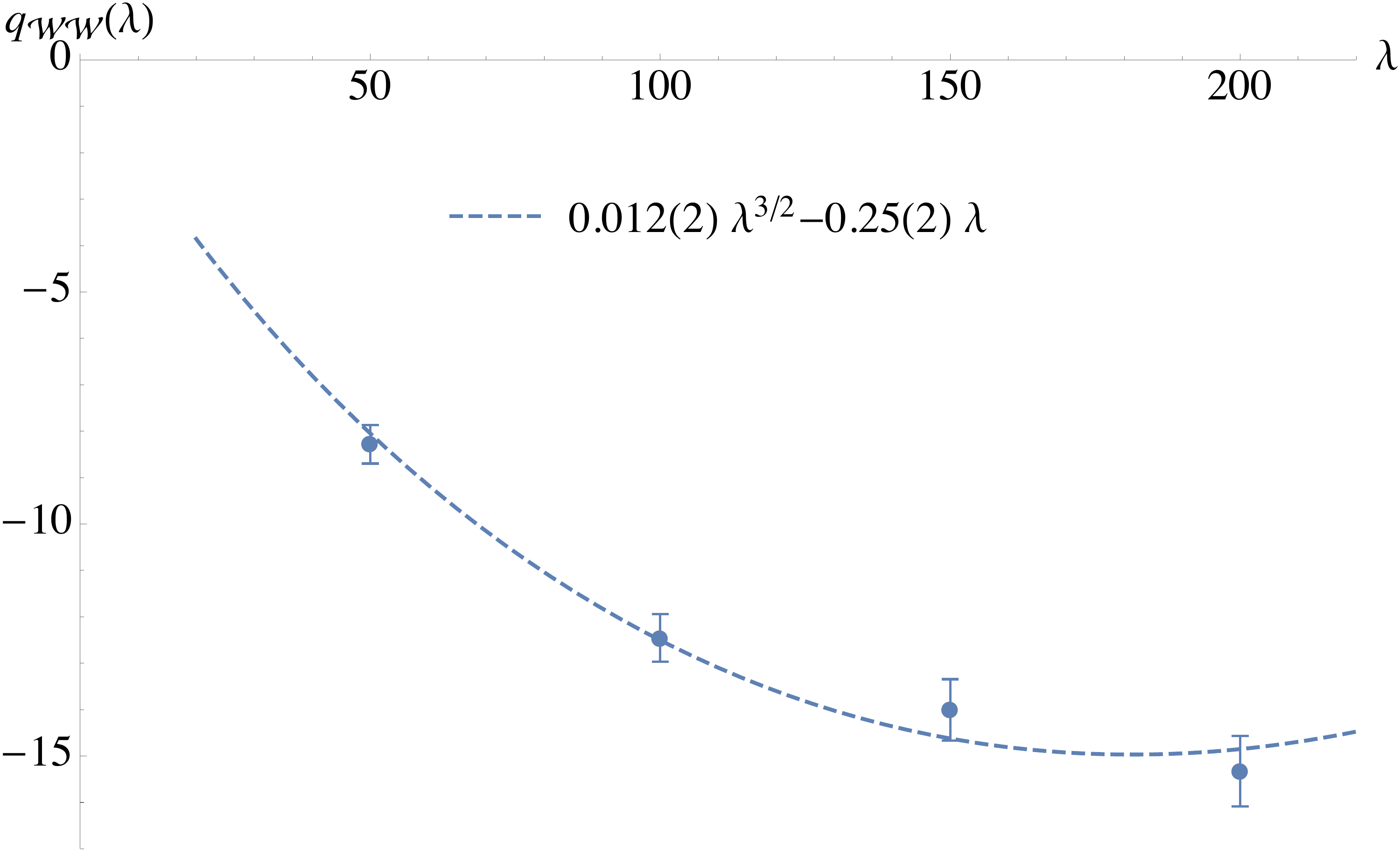}
\end{center}
\caption{ \label{fig:W1W2}\small 
Data for the $q_{_{\mc W\mc W}}(\lambda)$ function in \rf{664}    controlling  the $1/N^{2}$ correction to $\vev{\mc W_{1}\mc W_{2}}$ in the orbifold theory.}
\end{figure}

%%%%%%%%%%%%%%%%%%%%%%%%%
Finally, motivated by the discussion of the possible role of the D3-brane solution of \cite{Drukker:2005kx} in the $SU(N)\times SU(N)$ orbifold theory (see Introduction),    %and to make contact with it, 
we numerically computed 
the expectation value $\vev{\mc W_{1}\mc W_{2}}$  of the  two $SU(N)$ Wilson loops \rf{0} and determined (using  the same  fitting procedure  as 
discussed above) the associated $q_{_{\mc W\mc W}}(\lambda)$ function defined as  in  (\ref{1.9})
\be
\la{664}
\frac{\vev{\mc W_{1}\mc W_{2}}}{\vev{\mc W}_{0}^{2}} = 1+\frac{1}{N^{2}}\,q_{_{\mc W\mc W}}(\lambda)+\mc O\Big(\frac{1}{N^{4}}\Big).
\ee
The corresponding data points are shown in Fig.~\ref{fig:W1W2}. 
They decrease to negative values with  rate slower than the one observed in $q^{\rm orb}(\lambda)$. %Actually 
A best fit of the form (\ref{x7})
with $\eta$ fixed at $\frac{3}{2}$ gives $C_{\mc W\mc W} = +0.012(2)$ and $a_{1_{ \mc W\mc W} }= -21(2)$. The coefficient $C_{\mc W\mc W}$  has the opposite sign to the one in    \rf{xx7}  and  is close to the SYM value $\frac{1}{96}\approx  0.010$
in \rf{1.10}. %  (its sign is  and has the same sign, unlike (\ref{8.3}). 
One possible interpretation of this result  is that the  ``diagonal''  correlator  $\vev{\mc W_{1}\mc W_{2}}$  
of the two Wilson loops   in the fundamental representation 
exhibits the  (at the  leading non-planar order)  the   strong coupling    behaviour  
which is expected from  the D3-brane description, 
 while other terms appearing in \rf{XXX} 
 % the branching $SU(2N)\to SU(N)\times SU(N)$ (and characterized by an increasing 
%asymmetry in the couplings ) 
are less important in the large $k$ limit.

%%%%%%%%%%%%%%%%%%%%%%%%%%%%%
\subsection{Non-symmetric quiver}

\begin{figure}
\begin{center}
\includegraphics[width=0.45\textwidth]{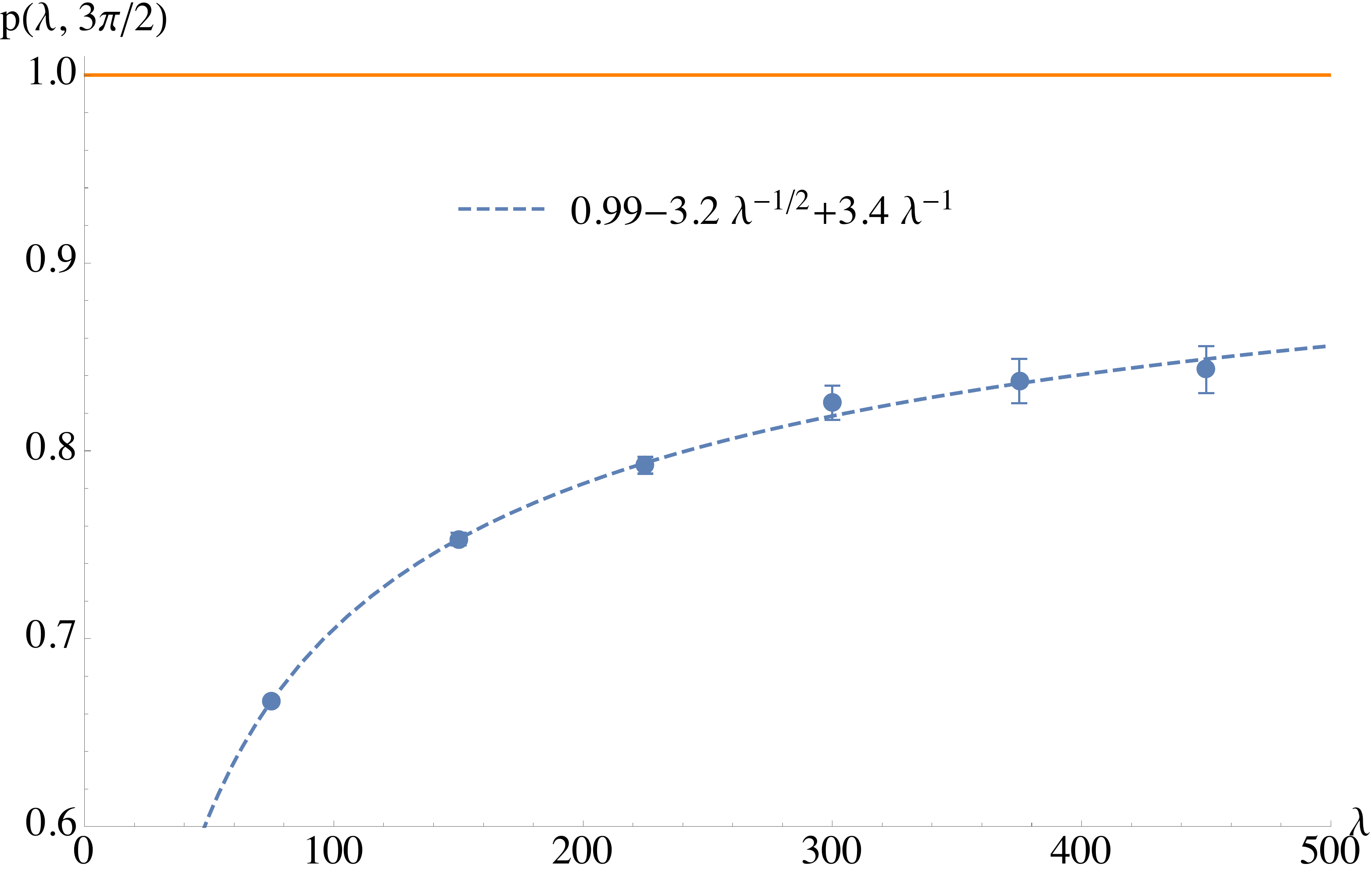}
\includegraphics[width=0.45\textwidth]{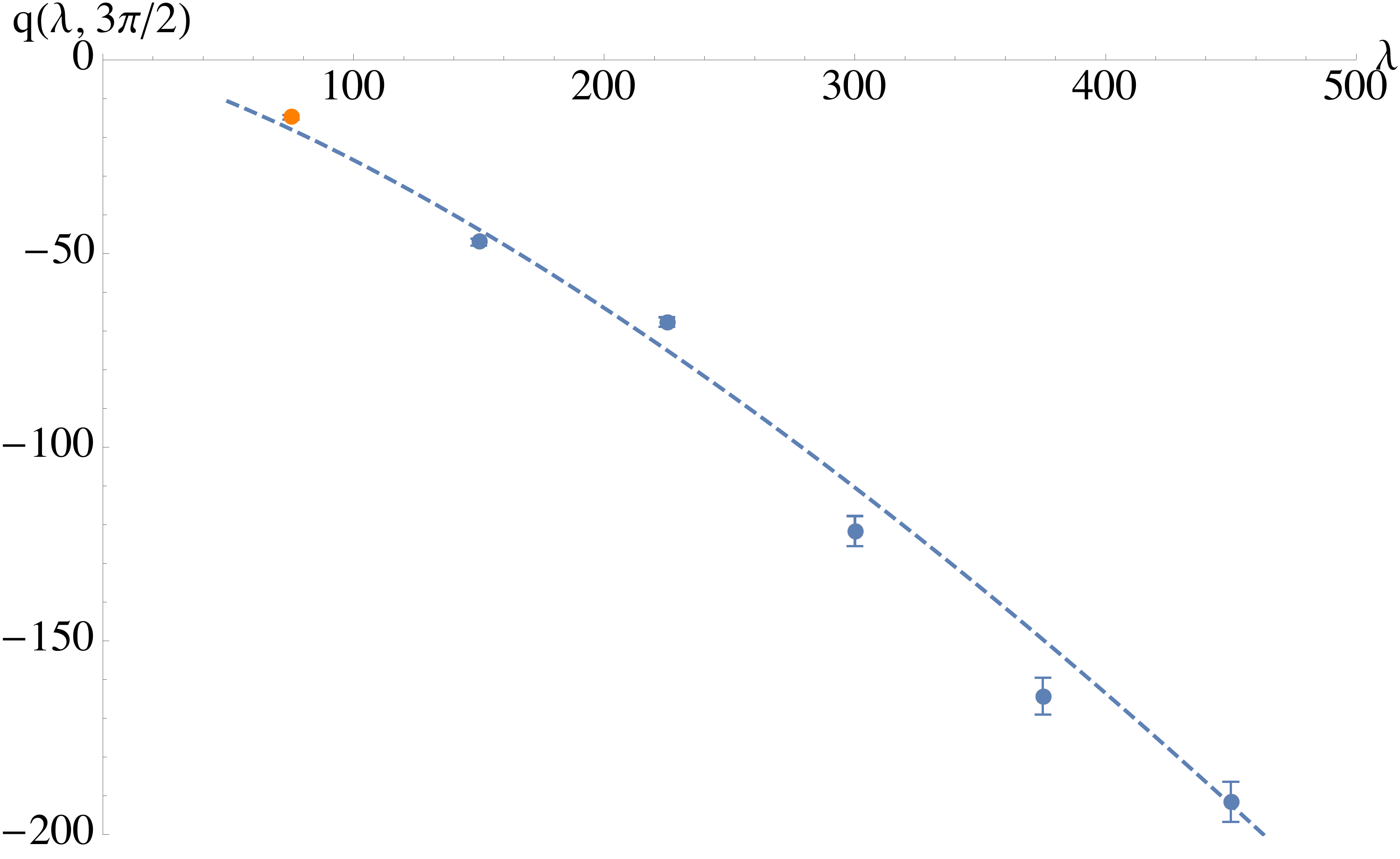}
\end{center}
\caption{ \label{fig:two-dual}\small 
Functions $p(\lambda, \frac{3\pi}{2})$ (left) and $q(\lambda, \frac{3\pi}{2})$ (right) for the quiver at the point $\lambda_{1}=3\lambda_{2}$,   with  $\lambda ={ 2\lambda_{1}\lambda_{2}\ov \lambda_{1}+\lambda_{2} }= {3\ov 2} \,\lambda_{2}$.
 The angle $\theta={3\pi \over  2}$  corresponds to the  Wilson loop for the second $SU(N)$ factor.}
%,  \ie the 
%one  computed with the eigenvalues of the $SU(N)$ gauge group element with coupling $\lambda_{2}$.}
\end{figure}

In the  case of generic (non-zero)   $\l_1$ and $\l_2$  the strong-coupling   asymptotics of the Wilson loops is given by \rf{2.10}. We   shall   study  the functions  $p(\l,\theta)$ 
  and $q(\l, \theta)$  in  the ratio \rf{125}  of $\vev{\mc W}_1$ to the planar SYM result. 
We begin with the special point  $\theta={\pi\ov 2}$ or  (see \rf{126})
\be\la{64}
\lambda_{2} = 3\,\lambda_{1}\, :\qquad 
 \lambda = \frac{3}{2}\lambda_{1}, \qquad 
w( \frac{\pi}{2}) = 2-   \frac{\pi}{2} = 0.429\dots\ .
\ee
The   numerical results are shown in Fig.~\ref{fig:two}. The left panel  gives  the function $p(\lambda,\frac{\pi}{2})$. As expected, 
it decreases  for large $\l$   towards 1  (this should hold for any $\theta$, see \rf{2.10})  and 
%A good fit, shown in the figure, is 
a good fit is 
\be
\la{ff1}
p(\lambda, {\pi\ov 2}) = 1.00+{0.23}\, {\lambda}^{1/2}+{8.2}\, {\lambda}^{-1} + ... \ .
\ee
Measurement of the  second Wilson loop   $\vev{\mc W}_2$   
provides  the information   about the same functions  at  the  complementary    value of the 
angle $\theta'= 2\pi-\theta = {3\pi\ov 2}$   for which  
\be
\lambda_{1} = 3\,\lambda_{2}\,:\qquad  \lambda = \frac{3}{2}\lambda_{1}, \qquad   \  w({3\pi\ov 2}) = 2+  {3\pi\ov 2} = 6.712\dots\ . 
\ee
%so that $w$  is quite larger than in \rf{64}. 
 The  corresponding results
 are shown in Fig.~\ref{fig:two-dual}. The best fit for  the $p(\lambda, {3\pi\ov 2})$ is\foot{The small but not negligible deviation of the estimated asymptotic value from 1  suggests that systematic errors should be reduced by
$N\to \infty$ extrapolations with larger values of $N$. This could be related to the much 
 large value of the correcting factor $w( {3\pi\ov 2})$ as compared to $w( {\pi\ov 2})$.} 
\be
\la{f2}
p(\lambda, {3\pi\ov 2}) = 0.99- {3.2}\, {\lambda}^{1/2} +{3.4}\, {\lambda}^{-1}+...\ . 
\ee
The function $q(\lambda; \theta)$ at $\theta={\pi\ov 2} $ and $ {3\pi\ov 2}$ is shown in the right panels of Fig.~\ref{fig:two} and Fig. \ref{fig:two-dual}. Our estimate for  the
exponent $\eta(\theta)$ in  the analog of (\ref{8.3}) is $\eta({\pi\ov 2}) = 1.3(2)$ and $\eta({3\pi\ov 2})=1.6(2)$. 
Both values  appear to be   similar to the one  found  in the orbifold case ($\theta= {\pi}$), i.e. 
 $\eta\approx  {3\ov 2} $. %, but the estimates have to be taken
%with some care due. 
It would be desirable to push the  MC  simulation to larger values of the coupling $\l$, but that 
 seems to require a dedicated analysis with a 
substantially increased computational power. 

\

\

%%%%%%%%%%%%%%%%%%%%%%%%%%%%%%%%%%%%%%
\section*{Acknowledgments}
%%%%%%%%%%%%%%%%%%%%%%%%%%%%%%%%%%%%%%
We would like to thank  N. Drukker, S. Giombi, J. Russo, 
%v2a
%E. Pomoni
and  K. Zarembo for  
 useful   discussions  and comments on the draft.
 %We also thank  B. Pioline for a  helpful communication. 
MB was supported by  the INFN grant GSS (Gauge Theories, Strings and Supergravity).
AAT was supported by the STFC grant  ST/T000791/1.   % ST/P000762/1.
%%%%%%%%%%%%%%%%%%%%%%%%%%%%%%%%
\begin{figure}
\begin{center}
\includegraphics[width=0.45\textwidth]{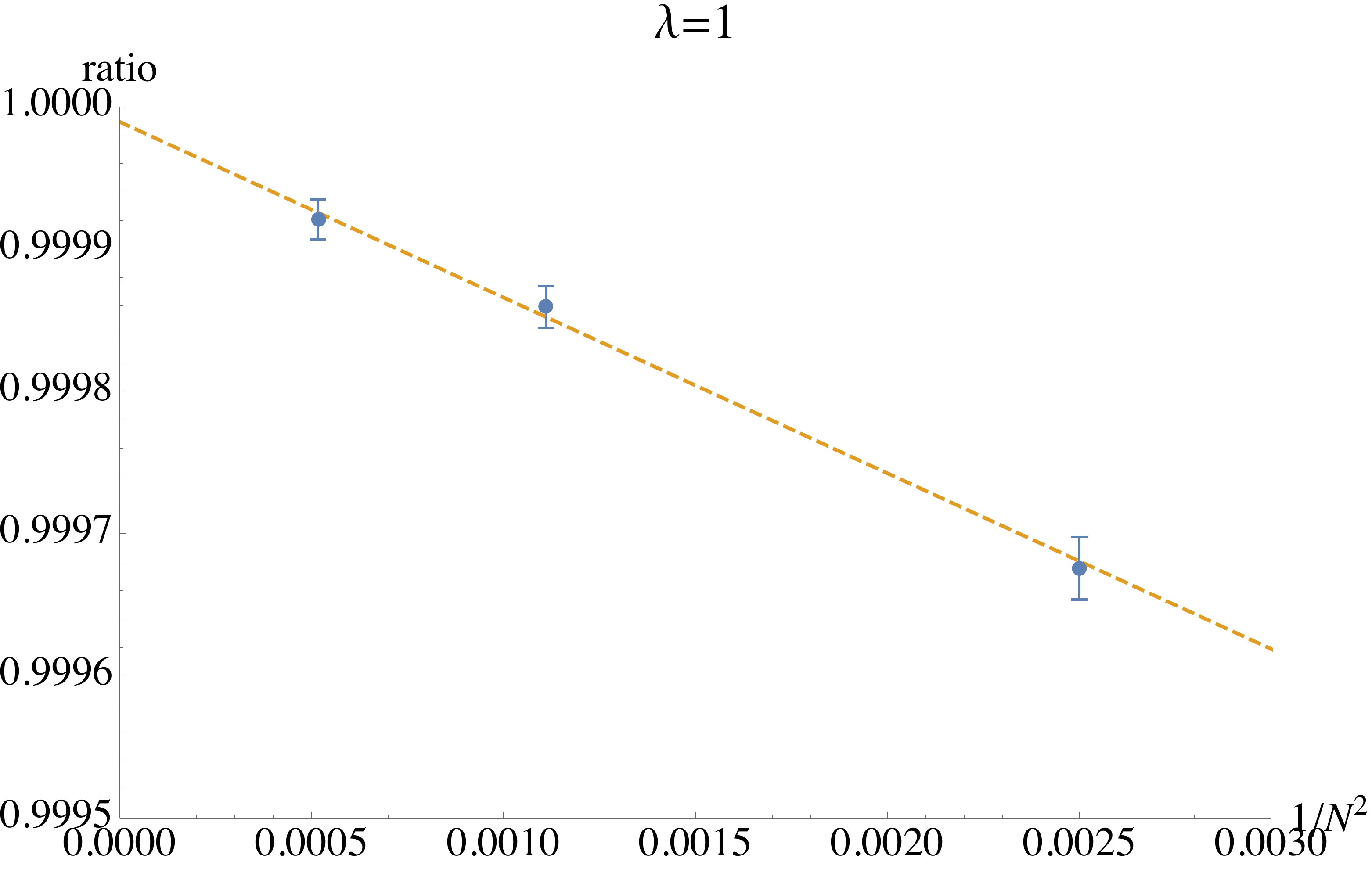}
\includegraphics[width=0.45\textwidth]{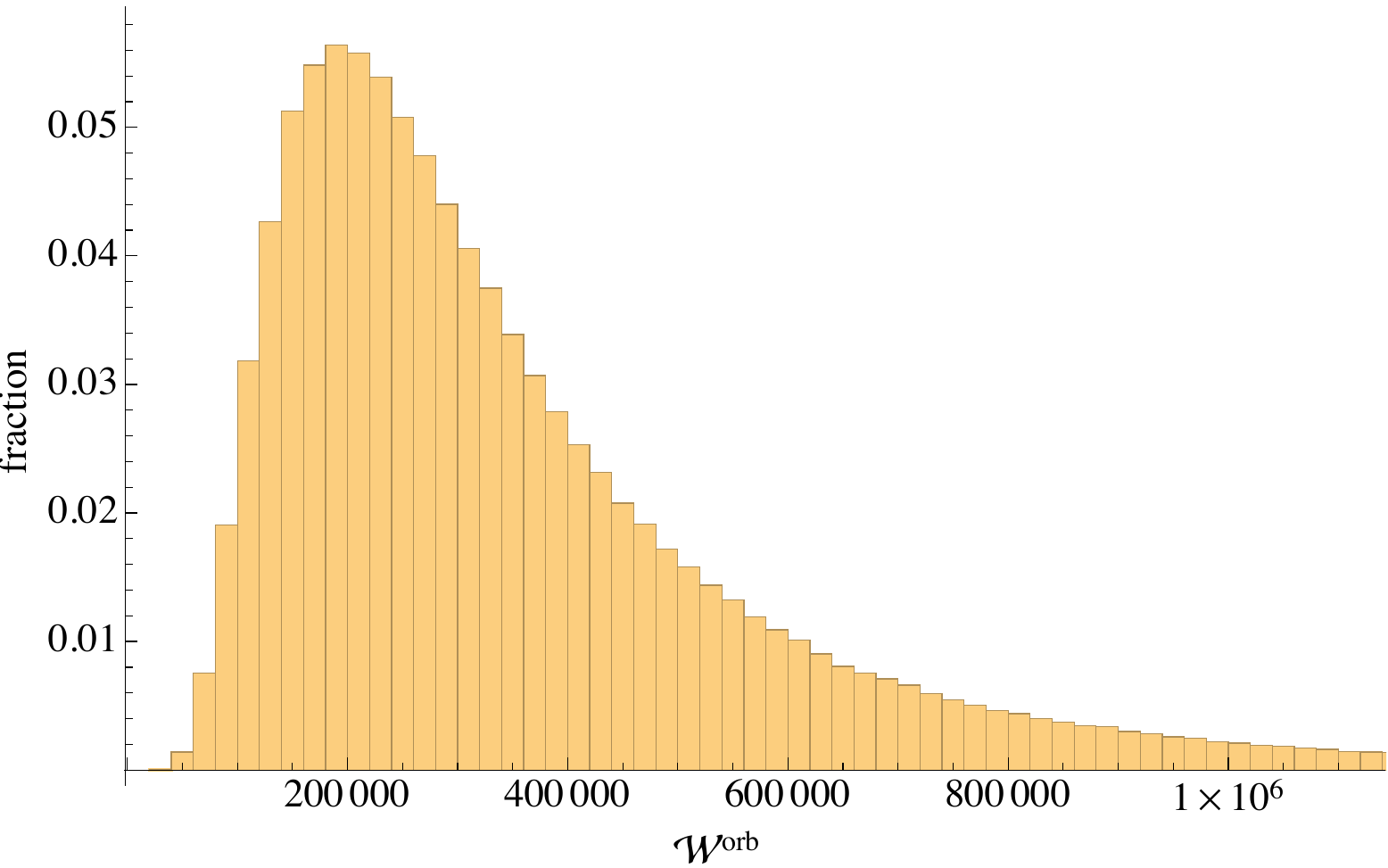}
\end{center}
\caption{\la{fig:Nfit} \small 
{\bf Left:} Fit of the ratio $\frac{{\vev{\mc W}}^{\rm orb} - {\vev{\mc W}}^{\rm SYM} }{\vev{\mc W}_0}$   (see \rf{1.1},(\ref{3.1}))
 with a linear function of  $1/N^{2}$ for $\l=1$. 
 The three data points correspond to $N=20,30,44$.  It is 
not necessary to take larger values since the intercept is already very close to the expected value  1. \ \
%The same behaviour can be observed using  the  exact finite $N,\lambda$ result for the 
%$SU(N)$ SYM theory in (\ref{3.17}).\ \ 
 {\bf Right}:   %Fit of the ratio $\frac{W(\lambda, N)}{\vev{\mc W}_0(\lambda)}$ 
 %with a linear function in $1/N^{2}$
% Similar    fit  for   $\lambda=400$;  the  data points correspond to $N=44,62,90$.
Histogram of the   Monte Carlo measurements of the orbifold Wilson loop from simulation at $\lambda=200$, $N=20$. For each (uncorrelated) Monte Carlo step, 
one records the measured value of $\mc W^{\rm orb}$ and the plot  shows the binned relative frequencies. The best statistical estimator for the quantum expectation value $\vev{\mc W}$ is the mean value of this empirical 
distribution.
}
\end{figure}
%%%%%%%%%%%%%%%%%%%%%%%%%%%%%%%%%%%%%%%%%%%
\begin{figure}
\begin{center}
\includegraphics[width=0.7\textwidth]{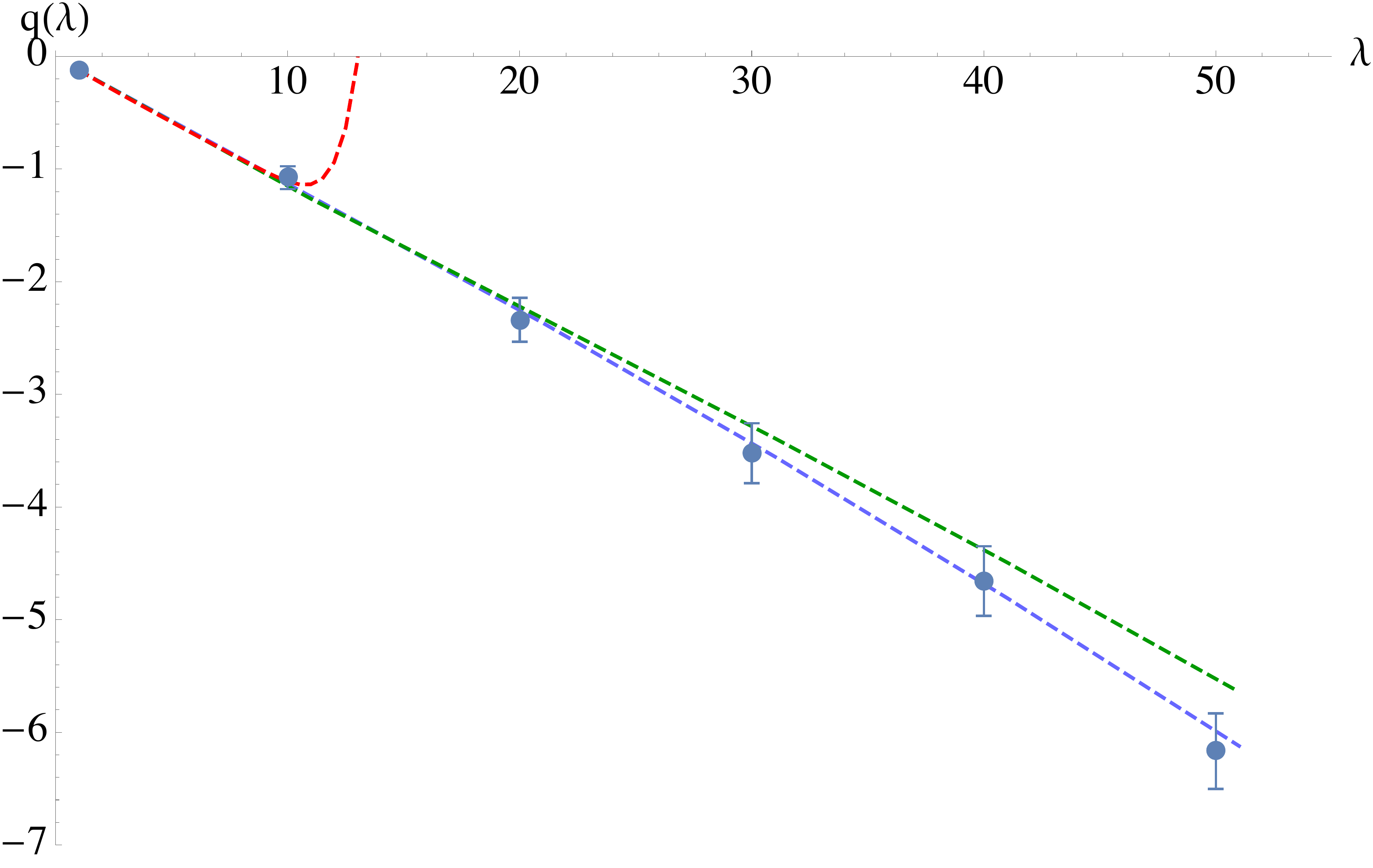}
\end{center}
\caption{
\la{fig:pade}\small 
Borel-Pad\'e resummation of the perturbative expansion of $q^{\rm orb}(\lambda)$. The red line is the perturbative expansion (\ref{322}), (\ref{329}) rapidly breaking 
down around $\lambda = \pi^{2}$. The green line is its [7/6] Pade' approximant already close to data, while the blue line is its numerical 
Borel transform.
}
\end{figure}
%%%%%%%%%%%%%%%%%%%%%%%%%%%%%%%%
\begin{figure}
\begin{center}
\includegraphics[width=0.45\textwidth]{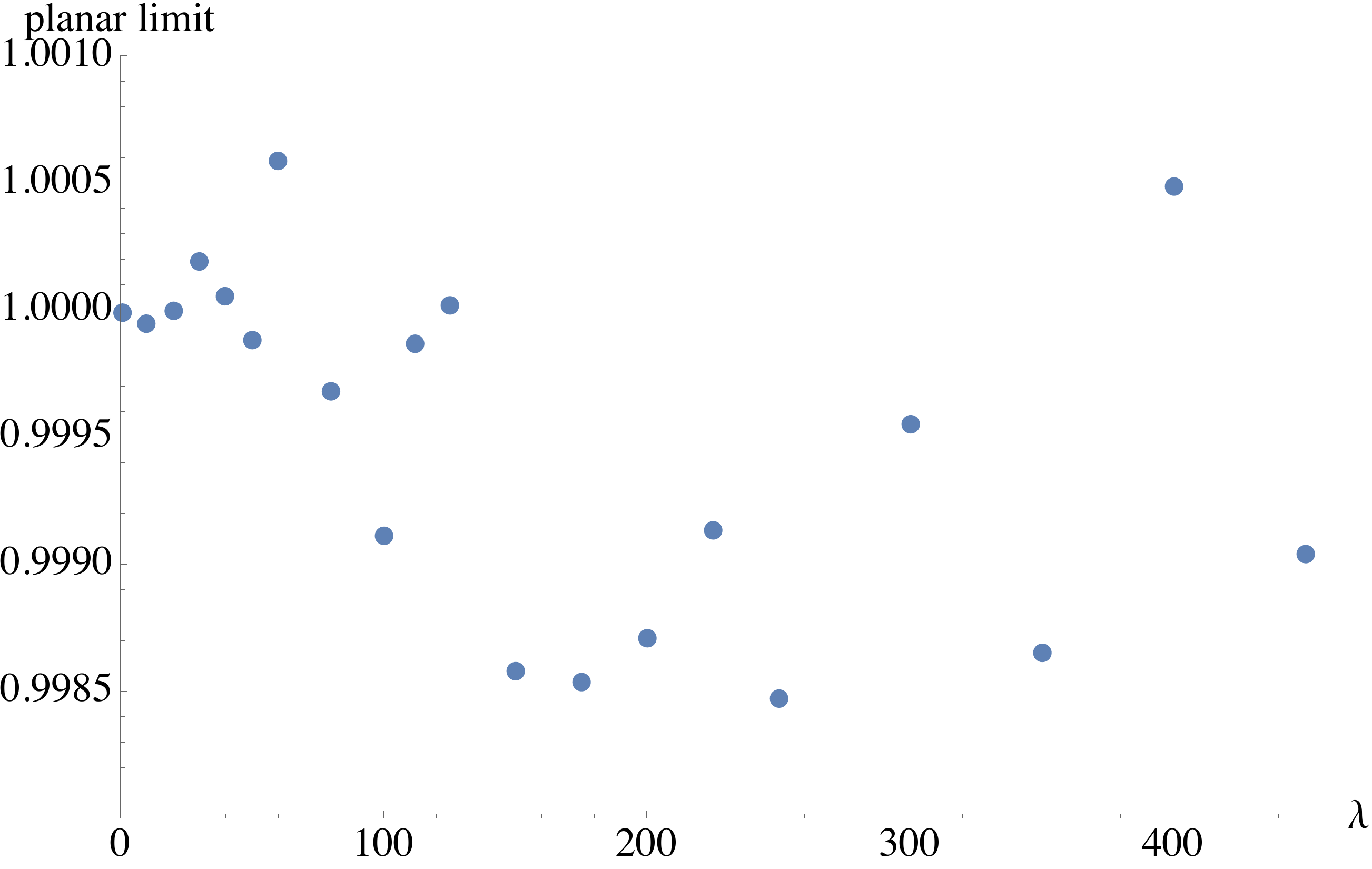}
\includegraphics[width=0.45\textwidth]{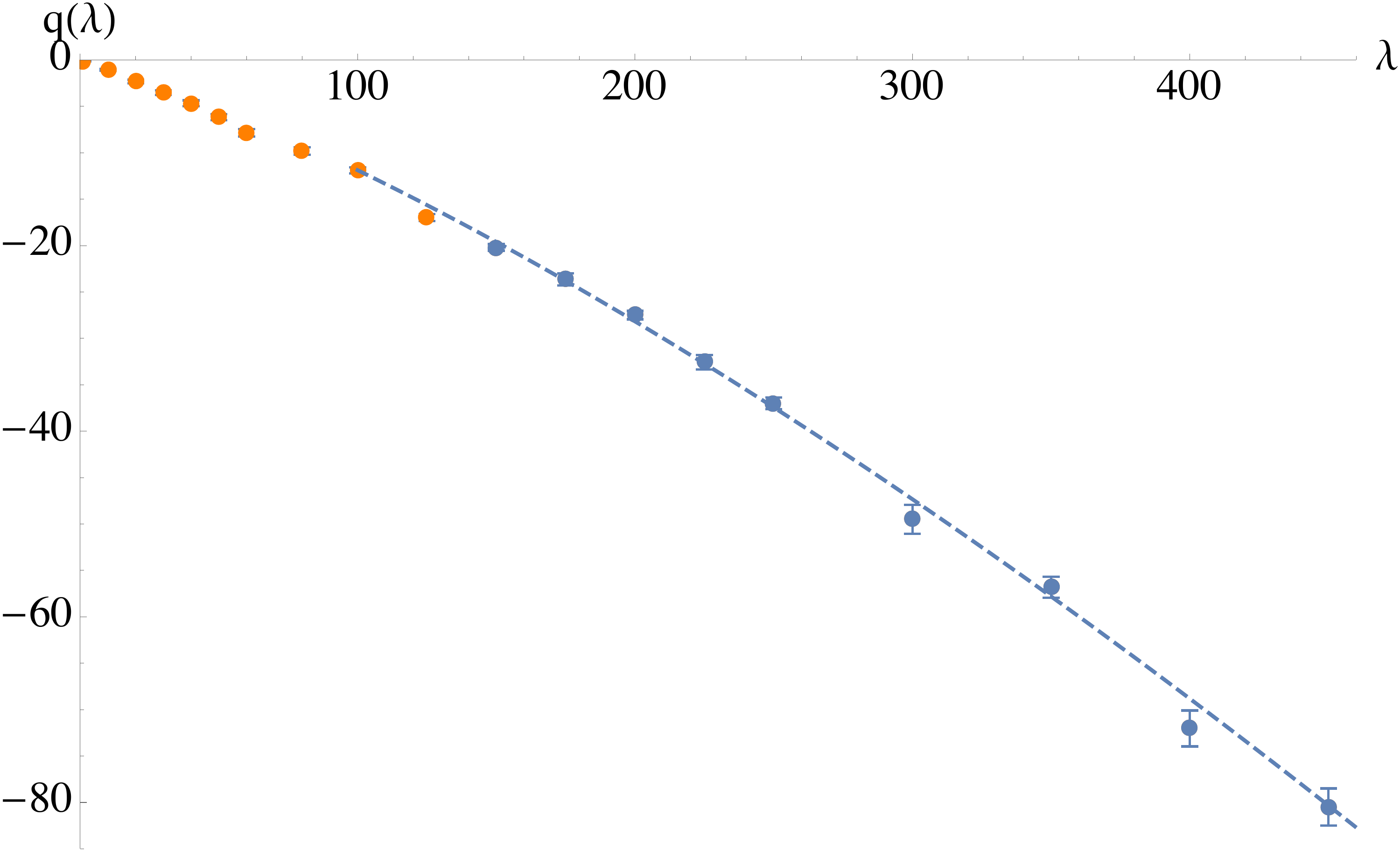}
\end{center}
\caption{\la{fig:res}\small 
{\bf Left}:\ 
Intercept in the large $N$ extrapolation %. According to (\ref{3.22}) it 
which should be equal to  1 due to the planar equivalence  %of the quiver theory
 with  the $\mc N=4$ SYM.
The deviation is a measure of the systematic error which can be reduced at the price of increasing the maximal 
 $N$ used in the simulations and in  the 
extrapolation to $N=\infty$.\ \ 
{\bf Right:} \ Data points for the  function $q^{\rm orb} (\lambda)$ defined in \rf{1.9},\rf{3.23}.  
 Dashed line is the non-linear fit with the functional
form $q^{\rm orb}(\lambda) = C\,\lambda^{\eta}(1+a_{1}\lambda^{-1/2} )$.
The fit is performed using data points with $\lambda \ge 100$ which have been determined with 
a relative error below 3\%\ .
}
\end{figure}
%%%%%%%%%%%%%%%%%%%%%%%%%%%%%%%%
\begin{figure}
\begin{center}
\includegraphics[width=0.45\textwidth]{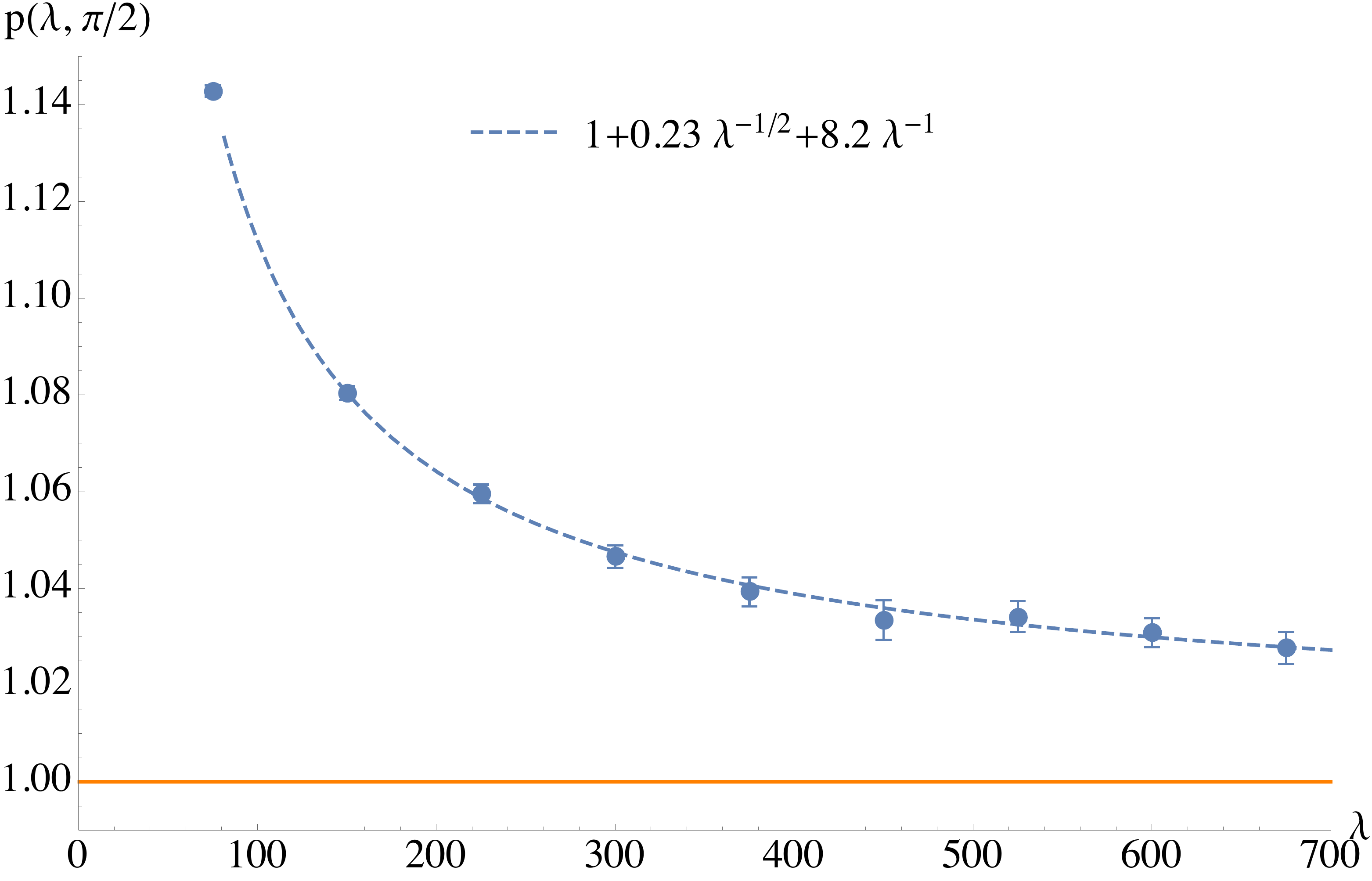}
\includegraphics[width=0.45\textwidth]{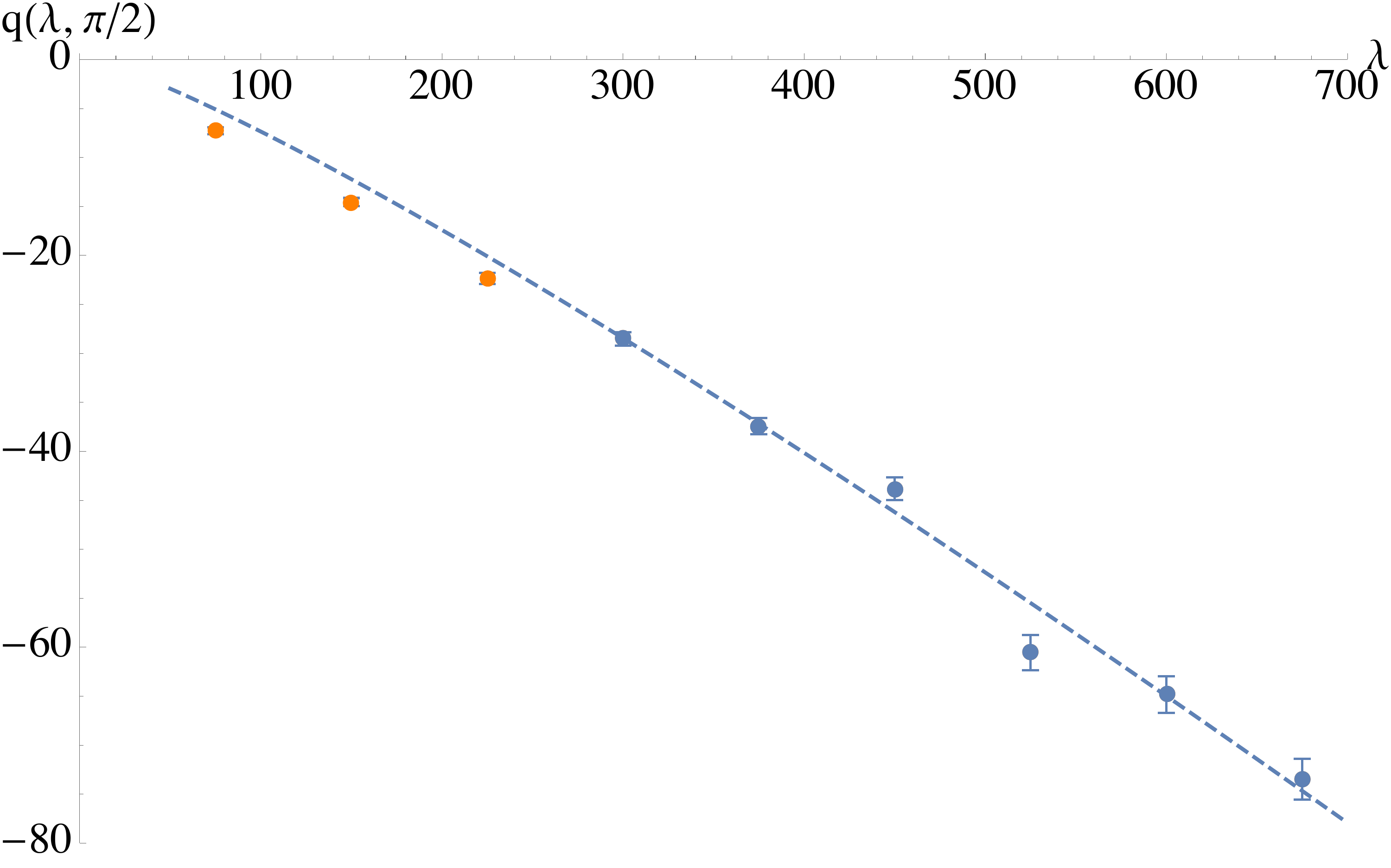}
\end{center}
\caption{
\la{fig:two}\small 
Functions $p(\lambda, \frac{\pi}{2})$ (left) and $q(\lambda, \frac{\pi}{2})$ (right) for the quiver at the point $\lambda_{2}=3\lambda_{1}$,  with   $\lambda = {2\lambda_{1}\lambda_{2}\ov  \lambda_{1}+\lambda_{2}}  = {3 \ov 2} \,\lambda_{1}$. 
}
\end{figure}
%%%%%%%%%%%%%%%%%%%%%%%%%%%%%%%%
%%%%%%%%%%%%%%%%%%%%%%%%%%%%%%%%

 \newpage 

%\section{Remarks}
%
%Can we use the methods in  \cite{Chen-Lin:2016kkk} ? 
%

%\section{$1/N$ expansion of general correlators}
%
%Is there a way to systematically compute the $1/N$ expansion of multi-trace correlators ? 
%
\def \no {\notag}

\appendix 
\section{Multi-trace $SU(N)$ recursion relations and $\vev{\mc W\,\mc O}_0$}
\la{app:rec}

The correlators $\vev{\mc W\, \mc O}_0 $  in a Gaussian  one-matrix model of the Wilson loop operator ${\mc W}= \tr\, 
e^{ \sqrt{ \l\ov 2 N} A}$ 
 and a  multi-trace chiral operator $\mc O$
may be reduced to a differential operator over  the 
coupling  constant  acting on $\vev{\mc W}_0$ (see \rf{88}--\rf{888}). This relation is exact at finite $N$ and is achieved by exploiting the
$SU(N)$ fusion/fission relations \cite{Billo:2018oog} and the associated recursion relations on the expectation values %of $T_{n_{1}, n_{2},...,n_r}$ in  \rf{36} 
\be
t_{n_{1}, n_{2},...,n_r} \equiv  \vev{\tr A^{n_{1}}\tr A^{n_{2}}\cdots\tr A^{n_r}}_0\ .
% \ , \qquad  T_{n_{1}, n_{2},...,n_r}=
%\tr A^{n_{1}}\tr A^{n_{2}}\cdots\tr A^{n_r}\ .
\ee
Let us consider as an example  $\vev{\mc W\,:\tr A^{6}:}_0$.  %From the explicit form of the normal ordered  $:\tr A^{6}: $
We find  ($g=\sqrt{ \l\ov N}$)  
\ba
&:\tr A^{6}: =  \tr A^{6}-3N \tr A^{4}+\frac{15}{4}N (\tr A)^{2}-3\tr A \tr A^{3}-\frac{3}{2}(\tr A)^{2}+\frac{15}{4}(N^{2}+1)\tr A^{2}-\frac{5}{8}(2N^{2}+N^{4}),\notag
\\
&\vev{\mc W\,:\tr A^{6}:}_0  =\sum_{k=0}^{\infty}\frac{g^{k}}{2^{\frac{k}{2}}\,k!}\te \Big(t_{k,6}- 3N t_{k,4}+\frac{15}{4}(N^{2}+1)t_{k,2}\notag\\  &\qquad \qquad \qquad\qquad \qquad\qquad\qquad \te 
-\frac{5}{8}(2N^{2}+N^{4})t_{k}+\frac{15}{4}N t_{k,1,1}-3\,t_{k,1,3}-\frac{3}{2}t_{k,2,2}
\Big).
\ea
Doing Wick contractions leads to a combination of ``single-trace'' terms that can be traded for  $\partial_g$ differential operators   
acting on $\vev{\mc W}_0$  and we  finally obtain 
\ba
 \vev{\mc W\,:\tr A^{6}:}_0 
&= \sum_{k=0}^{\infty}\frac{g^{k}}{2^{\frac{k}{2}}\,k!}\te \Big[
\frac{k}{2}t_{k+4}-k N t_{k+2}-\frac{3}{8}k(-1+2k-N^{2})t_{k}+\frac{3}{16}k(k-1) N t_{k-2}
\Big]\no  \\
&=  \sum_{k=0}^{\infty}\frac{g^{k}}{2^{\frac{k}{2}}\,k!}\te \Big[
\frac{2k(k-1)(k-2)(k-3)(k-4)}{g^{4}}-\frac{3}{8}k(-1+2k-N^{2})-\frac{2N k(k-1)(k-2)}{g^{2}}+\frac{3g^{2}N}{32}
\Big]\,t_{k}\notag
\\
  &= \Big[
2g\partial_{g}^{5}+\frac{3}{8}(N^{2}+1)\,g\partial_{g}-\frac{3}{4}(g\partial_{g})^{2}-2N\, g\partial_{g}^{3}+\frac{3N}{32}g^{2}
\Big]\,\vev{\mc W}\ .
\ea
This procedure can be easily coded in symbolic manipulation programs.

%%%%%%%%%%%%%%%%%%%%%%%%%%%%%%%%%%%%%
\section{Coefficient functions of    $\zeta$-terms  in $\vev{\mc W}^{\rm orb}$}
\la{B}

Here we   shall   provide some details  of  the weak-coupling  computation of the  coefficient functions $W_\zeta$ in \rf{3.1}. 
Generalizing the calculation in (\ref{327}), the contribution proportional to a single $\zeta_{2n+1}$  to the expectation value $\vev{f}_0$ is  
given by 
{\small 
\ba
\vev{f}_0 &= 1+2\sum_{n=1}^{\infty}\Big(\frac{\lambda}{8\pi^{2}N}\Big)^{n+1}\frac{(-1)^{n}}{n+1}\zeta_{2n+1}\sum_{k=0}^{2n+2}
(-1)^{k}\binom{2n+2}{k}\vev{\tr A_{1}^{k}\,\tr A_{1}^{2n+2-k}}_{0, \rm c} +\mc O(\zeta^{2}) \notag \\
&= 1+2\sum_{n=1}^{\infty}\Big(\frac{\lambda}{8\pi^{2}N}\Big)^{n+1}\frac{(-1)^{n}}{n+1}\zeta_{2n+1}\sum_{k=0}^{n+1}
\binom{2n+2}{2k}\vev{\tr A_{1}^{2k}\,\tr A_{1}^{2(n-k+1)}}_{0,\rm  c}\notag \\
&\ \ \  \ \ \  -2\sum_{n=1}^{\infty}\Big(\frac{\lambda}{8\pi^{2}N}\Big)^{n+1}\frac{(-1)^{n}}{n+1}\zeta_{2n+1}\sum_{k=0}^{n}
\binom{2n+2}{2k+1}\vev{\tr A_{1}^{2k+1}\,\tr A_{1}^{2n-2k+1}}_{0,\rm  c}+O(\zeta^{2}).
\ea
}
Using that  the connected  correlators are  given by  \cite{Beccaria:2020hgy}
\ba
\la{5.7}
\vev{\tr A^{2k_{1}}\tr A^{2k_{2}}}_{0,\rm c} &= N^{k_{1}+k_{2}}\frac{2^{k_{1}+k_{2}}\,\Gamma(k_{1}+\frac{1}{2})\Gamma(k_{2}+\frac{1}{2})}{\pi\,(k_{1}+k_{2})\,\Gamma(k_{1})\Gamma(k_{2})} +\mc O(N^{k_{1}+k_{2}-2}), \notag \\
\vev{\tr A^{2k_{1}+1}\tr A^{2k_{2}+1}}_{0,\rm c} &= N^{k_{1}+k_{2}+1}\frac{2^{k_{1}+k_{2}+1}\,k_{1}\,k_{2}\,\Gamma(k_{1}+\frac{3}{2})\Gamma(k_{2}+\frac{3}{2})}{\pi\,(k_{1}+k_{2}+1)\,\Gamma(k_{1}+2)\Gamma(k_{2}+2)} +\mc O(N^{k_{1}+k_{2}-1}), 
\ea
%v2
we get
{\small 
\ba
\vev{f}_0 &\stackrel{N\to \infty}{=}1+2\sum_{n=1}^{\infty}\Big(\frac{\lambda}{8\pi^{2}}\Big)^{n+1}\frac{(-1)^{n}}{n+1}\zeta_{2n+1}\sum_{k=0}^{n+1}
\binom{2n+2}{2k}\frac{2^{n+1}\,\Gamma(k+\frac{1}{2})\Gamma(n+1-k+\frac{1}{2})}{\pi\,(n+1)\,\Gamma(k)\Gamma(n+1-k)}\notag \\
&-2\sum_{n=1}^{\infty}\Big(\frac{\lambda}{8\pi^{2}}\Big)^{n+1}\frac{(-1)^{n}}{n+1}\zeta_{2n+1}\sum_{k=0}^{n}
\binom{2n+2}{2k+1}\frac{2^{n+1}\,k(n-k)\Gamma(k+\frac{3}{2})\Gamma(n-k+\frac{3}{2})}{\pi\,(n+1)\,\Gamma(k+2)\Gamma(n-k+2)}+\mc O(\zeta^{2})\notag \\
%%%%
%&= 1+2\sum_{n=1}^{\infty}\Big(\frac{\lambda}{8\pi^{2}}\Big)^{n+1}\frac{(-1)^{n}}{n+1}\zeta_{2n+1}
%2^{3n+1}\bigg[\frac{\Gamma(n+\frac{1}{2})\Gamma(n+\frac{3}{2})}{\pi\,\Gamma(n)\Gamma(n+2)}-
%\Gamma(n+\frac{1}{2})\Gamma(n+\frac{3}{2})}{\pi\Gamma(n-1)\Gamma(n+3)}\bigg]
%+\mc O(\zeta^{2}) \notag \\
&= 1+2\sum_{n=1}^{\infty}\Big(\frac{\lambda}{8\pi^{2}}\Big)^{n+1}\frac{(-1)^{n}}{n+1}\zeta_{2n+1}\frac{3}{\pi}\frac{2^{3n+1}\Gamma(n+\frac{1}{2})\Gamma(n+\frac{3}{2})}{\pi\Gamma(n)\Gamma(n+3)}+\mc O(\zeta^{2}) \ . 
\ea
}
This  leads to (\ref{42})  after  using (\ref{4.4}).

To prove the relation \rf{411}  for the contribution to $\Delta q$  of the sum of  terms  proportional to powers of $\z_3$
one may start with   the following   $U(N)$ 2-matrix $(A,B)$    model with the $\zeta_{3}$ term  in the exponent 
representing the corresponding contribution  coming from  $f$  in \rf{2.1},\rf{3.7}\foot{In this special case there will be no difference between $U(N)$ and $SU(N)$ cases.}
\be\la{b14}
\Z=\lim_{N\to \infty} \Z_N\ , \qquad \Z_N(\xi) = \int [dA dB] \ \ e^{-\tr A^{2}-\tr B^{2} -\frac{\xi}{N^{2}}(\tr A^{2}-\tr B^{2})^{2}} ,\qquad
 \xi\equiv 3\,\zeta_{3}\,\Big(\frac{\lambda}{8\pi^{2}}\Big)^{2}.
\ee
Then according to (\ref{4.4}),
\be
\la{5.29}
\Delta q(\lambda)\Big|_{\sum_n \zeta^n_{3}}  = \frac{\lambda^{2}}{8}\frac{d}{d\lambda}\log \Z \ . 
\ee
Since the integrand in $\Z_N$ depends only on $\tr A^{2}$ and $\tr B^{2}$,  introducing the  radial coordinates
$r_{A}, r_{B}$  we get (ignoring irrelevant constant factor) 
\be
\Z_{N}(x) = \int_{0}^{\infty} dr_{A}\, dr_{B}\, r_{A}^{N-1+N(N-1)}\,  r_{B}^{N-1+N(N-1)}
\ e^{-r_{A}^{2}-r_{B}^{2} -\frac{\xi}{N^{2}}(r_{A}^{2}-r_{B}^{2})^{2}}\ . 
\ee
The large $N$ limit  is  found from  a  saddle  point of  the effective action 
$
S_{\rm eff}= (N^{2}-1)\log(r_{A}r_{B})-\frac{\xi}{N^{2}}(r_{A}^{2}-r_{B}^{2})^{2}-r_{A}^{2}-r_{B}^{2}.
$
Choosing the symmetric saddle with $r_{A}=r_{B}=\sqrt\frac{N^{2}-1}{2}$
and integrating over the  fluctuations gives
\be\la{b17}
\Z(\xi) = \frac{1}{\sqrt{1+2\xi}}\ .
\ee
As a result, using \rf{5.29} we  find  the strong-coupling asymptotics in \rf{411}. 
An alternative more rigorous and general  approach is based on observing that $\Z$  in \rf{b14}  may be represented as
\ba
&\Z(\xi)= \lim_{N\to \infty}e^{-\xi(\partial^2_{x}+\partial^2_{y}-2\partial^2_{xy})}Z(x)Z(y)\Big|_{x=y=0}\ ,\qquad \notag \\
& Z(x) \equiv   \int [dA]\ e^{-\tr A^{2}+\frac{x}{N}\tr A^{2}} = \Big(1-\frac{x}{N}\Big)^{-\frac{N^{2}-1}{2}} = 
 e^{-\frac{N}{2}x+\frac{1}{4}x^{2}+\mc O(1/N)}\ .\la{b19}
\ea
As a result, we  get   again   \rf{b17}. 

Similar approach can be used  to derive \rf{415} for the contribution of 
 terms  proportional to  products of $\z_3, \z_5, ...,\z_{2k+1}$. 
For example,   let us  consider the $\z_3\z_5$  terms. The new interaction term in the exponent in the analog of  \rf{b14} 
will be 
\ba
&\Delta S_{\z_5}= -\frac{\eta}{N^{3}}\Big [2(\tr A^{3})^{2}+2(\tr B^{3})^{2}-3\tr A^{2}\tr A^{4}-3 \tr B^{2}\tr B^{4}\no \\  &
\qquad \qquad \qquad  +3 \tr A^{4}\tr B^{2}+3 \tr A^{2}\tr B^{4}-4\tr A^{3}\tr B^{3}\Big ]  \ , \qquad \qquad 
\eta  = -\frac{10}{3}\zeta_{5}\Big(\frac{\lambda}{8\pi^{2}}\Big)^{3}\ .
\ea
In this case  instead   of  \rf{b19} we will need to consider 
\ba
\la{6.10}
\Z(\xi, \eta) = &\lim_{N\to \infty} \exp\Big[-\xi(\partial^2_{x_{1}}+\partial^2_{y_1}-2\partial^2_{x_1 y_1 })
 -\eta (2 \partial^2_{x_2}+2\partial^2_{y_2}-3\partial^2_{x_1 x_3 }-3\partial^2_{y_1 y_3 } \lp\qquad \qquad 
+3 \partial^2_{x_3 y_1}+3\partial^2_{x_1 y_3}
-4\partial^2_{x_2 y_2})
\Big]\,Z(x_1 , x_2, x_3 )\ Z(y_1,y_2,y_3)\Big|_{x_{i}=y_{i}=0},
\\ 
Z(x_1,x_2,x_3) = &\int [dA]\ e^{-\tr A^{2}+\frac{x_1}{N}\tr A^{2}+\frac{x_2}{N^{3/2}}\tr A^{3}+\frac{x_3}{N^{2}}\tr A^{4}} \ . \la{b11}
\ea
Expanding  \rf{b11}, taking log    and sending $N\to\infty$  we find  
 %gives % the remarkable simple structure
\be
Z(x_1,x_2,x_3) = \exp\Big[\frac{N}{2}(x_1 + x_3)+\frac{1}{4}x_1^{2}+\frac{3}{16}x_2^{2}+x_1 x_3 +\frac{9}{8}x_3^{2}+\mc O(1/N)\Big]\ . 
\ee
Using this in \rf{6.10}  gives
\ba
\Z(\xi, \eta)  = &\te 1-3 \zeta_{3} \Big(\frac{\lambda}{8\pi^2}\Big)^2+15 \zeta_{5} \Big(\frac{\lambda}{8\pi^2}\Big)^3+\frac{27 \zeta_{3}^2 }{2}\Big(\frac{\lambda}{8\pi^2}\Big)^4-165 
\zeta_{3} \zeta_{5} \Big(\frac{\lambda}{8\pi^2}\Big)^5 \lp\te 
+\Big(-\frac{135 \zeta_{3}^3}{2}+\frac{1125 
\zeta_{5}^2}{2}\Big) \Big(\frac{\lambda}{8\pi^2}\Big)^6+\frac{2565}{2} \zeta_{3}^2 \zeta_{5} \Big(\frac{\lambda}{8\pi^2}\Big)^7+\dots
\ea
As a result
\ba
\Delta q(\lambda)\Big|_{\zeta_{3}, \zeta_{5}} = & \frac{\lambda^{2}}{8}\frac{d}{d\lambda}\log \Z
= \te -\frac{3 \zeta_{3} }{4}\lambda \Big(\frac{\lambda}{8\pi^2}\Big)^{2}+\frac{45 \zeta_{5} }{8}\lambda \Big(\frac{\lambda}{8\pi^2}\Big)^{3}+\frac{9 \zeta 
(3)^2 }{2}\lambda \Big(\frac{\lambda}{8\pi^2}\Big)^{4} -75 \zeta_{3} \zeta_{5} \lambda \Big(\frac{\lambda}{8\pi^2}\Big)^5\lp \te \qquad \qquad \qquad \quad 
 +\Big(-27 \zeta_{3}^3+\frac{675 
\zeta_{5}^2}{2}\Big) \lambda \Big(\frac{\lambda}{8\pi^2}\Big)^6+630 \zeta_{3}^2 \zeta_{5} \lambda \Big(\frac{\lambda}{8\pi^2}\Big)^7+\cdots\ , \la{b144} 
\ea
which agrees with the results  given  in the main text.

It is interesting to note  that (\ref{6.10}) can be computed in  a closed form
generalizing \rf{b17}  %.  Guided from the $\xi'=0$ case, we check that 
\be\la{b15} 
\Z(\xi, \eta) = \Big[1+2\xi-9\eta+6\xi\eta-\frac{81}{2}\eta^{2}-\frac{27}{2}\eta^{3}\Big]^{-1/2}.
\ee
Applying $ \frac{\lambda^{2}}{8}\frac{d}{d\lambda}$  to the log of  \rf{b15}  as in \rf{b144}  then gives the exact form of 
$\Delta q(\lambda)\Big|_{\zeta_{3}, \zeta_{5}} $.

%%%%%%%%%%%%%

%%%%%%%%%%%%%

\section{Wilson loop  in   $SU(N)$  ``orientifold''    $\mc N=2$   superconformal  theory  } %IIB Orientifolds  with weak planar equivalence}

It is possible to give a similar discussion of  the  large $N$ expansion  of  the Wilson loop $\vev{\mc W}$ and the free energy 
in a particular $\mc N=2$   superconformal   gauge  theory involving  in addition to the  $SU(N)$  $\N=2$ vector multiplet  also 
two  hypermultiplets --  in  rank-2 symmetric  and antisymmetric  $SU(N)$ representations. 
This theory   admits a regular 't Hooft large $N$  limit   and thus  is similar to the quiver  theory discussed   above.  It 
    should be dual to the type IIB superstring 
on a particular  orientifold 
AdS$_{5}\times S^{5}/(\mathbb Z_{2}^{\rm orient}\times \mathbb Z_{2}^{\rm orb})$
 (see \cite{Ennes:2000fu}). 
%The second model $M_{2}$ has a more intricate orientifold/orbifold structure and reduces in near horizon limit to $AdS_{5}\times %S^{5}/\mathbb Z_{4}$, where $\mathbb Z_{4}$ mixes the $\mathbb Z_{2}$ orbifold and orientifold twist.
%In both cases, we are dealing with a discrete quotient of $S^{5}$ and this will be responsible for a weak form of 
%planar equivalence with $\mc N=4$ SYM.
 
This  theory  is one of the  five cases of $\mc N=2$ superconformal theories
admitting a gauge group $SU(N)$ with generic $N$ \cite{Koh:1983ir}.  The corresponding 
     BPS  circular Wilson loop is  again equal
to the $\mc N=4$ SYM one at the planar level.\foot{This planar equivalence extends to classes of  ``even'' observables, while ``odd'' sectors display  deviations from SYM  case already at  the planar level  \cite{Beccaria:2020hgy}.}
Here we shall focus  on  the weak-coupling expansion of the  first subleading $1/N^2$   correction, i.e.  of 
the  corresponding  function $q(\l)$ defined  as in \rf{1.9}. 

From  the supersymmetric localization, the free  energy and the Wilson  loop expectation value  $\vev{\mc W}^{\rm orient}$
in this theory   are   described by 
the Hermitian one-matrix  model of the similar structure as in \rf{2.1}  where instead of \rf{3.7}  now   we have 
\cite{Billo:2019fbi}
\ba
\log f &= 2\,\sum_{n=1}^{\infty}(-1)^{n+1}\Big(\frac{\lambda}{8\pi^{2}N}\Big)^{n+1}\frac{\zeta(2n+1)}{n+1}\sum_{k=1}^{n-1}
\binom{2n+2}{2k+1}\,\tr A^{2k+1}\tr A^{2n-2k+1} \ .  \la{c1}
\ea
One can then organise  the expansion of $\vev{\mc W}^{\rm orient}$  in powers of monomials of $\zeta_{2n+1} $-constants as in \rf{3.1}. 
One finds that, as in the orbifold theory, at the leading non-planar level  all appearing $\zeta_{2n+1}$-monomials       
  are multiplied by  $I_{1}(\sqrt\lambda)$ times a power of $\lambda$ (cf. \rf{4.1}).
  Explicitly,  for $\Delta q$ defined as in \rf{3.23}, i.e.  
  $\Delta q=  q^{\rm orient}-q^{\rm SYM}$,    we find 
%%%%%%%%%
\ba
\frac{1}{8\pi^{2}} & \Delta q(\lambda) =\te  -\frac{15 \zeta _5}{4}\,\big(\frac{\lambda}{8\pi^2}\big)^{4}+\frac{105 \zeta _7}{2}\,\big(\frac{\lambda}{8\pi^2}\big)^{5}
-\frac{2205 \zeta _9}{4}\,\big(\frac{\lambda}{8\pi^2}\big)^{6}+\big(\frac{75 \zeta _5^2}{2}+\frac{10395 \zeta 
_{11}}{2}\big) \,\big(\frac{\lambda}{8\pi^2}\big)^7  \lp\te 
+\big(-\frac{3675}{4} \zeta _5 \zeta _7-\frac{1486485 \zeta
_{13}}{32}\big) \,\big(\frac{\lambda}{8\pi^2}\big)^8+\big(\frac{22785 \zeta _7^2}{4}+8505 \zeta _5 \zeta 
_9+\frac{6441435 \zeta _{15}}{16}\big)\, \big(\frac{\lambda}{8\pi^2}\big)^9  \lp\te 
+\big(-375 \zeta _5^3-\frac{853335 
\zeta _7 \zeta _9}{8}-\frac{571725 \zeta _5 \zeta 
_{11}}{8}-\frac{109504395 \zeta _{17}}{32}\big)\, \big(\frac{\lambda}{8\pi^2}\big)^{10}  \lp\te 
+\big(13125 \zeta _5^2 
\zeta _7+504630 \zeta _9^2+\frac{3620925 \zeta _7 \zeta 
_{11}}{4}+\frac{4601025 \zeta _5 \zeta _{13}}{8}+\frac{459349605 
\zeta _{19}}{16}\big) \,\big(\frac{\lambda}{8\pi^2}\big)^{11}   + ...\ . 
\ea
%%%%%%%%%%%%
%\lp +\Big(-\frac{2465925}{16} \zeta _5 \zeta _7^2-\frac{467775}{4} \zeta _5^2 \zeta _9-\frac{276600555 \zeta _9 
%\zeta _{11}}{32}-\frac{470224755 \zeta _7 \zeta _{13}}{64}-\frac{289864575 \zeta _5 \zeta _{15}}{64}  \lp
%-\frac{61093497465 \zeta _{21}}{256}\Big) \,\Big(\frac{\lambda}{8\pi^2}\Big)^{12} +\cdots.
%\ea
Like in \rf{x2},\rf{1.16}  there is again a relation   between $\Delta q$ and the large $N$ limit of the difference of the orientifold  and 
$SU(N)$  SYM  free energies
%\foot{Here the planar equivalence is to one copy of $SU(N)$ SYM  so there is no factor of 2 
%in front of $F^{\rm SYM}$. The presence of one factor of $SU(N)$ is also the reason for a different coefficient in 
%front of $\frac{d}{d\lambda}$   compared to \rf{x2}. } 
\foot{Note that as  both the $\N=2$  orientifold   theory  and  the $\N=4$ SYM  theory here  
 are defined for  a single copy of $SU(N)$  
the coefficients in \rf{c33}    are different from those in \rf{x2}   by  factors of 2.}
\ba
\Delta q(\lambda) = - \frac{\lambda^{2}}{4}\frac{d}{d\lambda}\,\Delta F(\lambda)\ , \qquad \qquad 
\Delta F(\lambda) =  \lim_{N\to \infty }\big[F^{\rm orient}(\lambda; N)- F^{\rm SYM}(\lambda; N)\big]\ . \la{c33}
\ea

\newpage

\bibliography{BT-Biblio}
\bibliographystyle{JHEP}
\end{document}